# Cloud Migration Process

# A Survey, Evaluation Framework, and Open Challenges

Mahdi Fahmideh, Graham Low, Ghassan Beydoun, Farhad Daneshgar

University of New South Wales, Sydney, Australia

**ABSTRACT:** Moving mission-oriented enterprise applications to cloud environments is a major IT strategic task and requires a systematic approach. The foci of this paper are to review and examine existing cloud migration approaches from the process models perspective. To this aim, an evaluation framework is proposed and used to analyse and compare existing approaches for highlighting their features, similarities, and key differences. The survey distills the state of the art in cloud migration research and makes a rich inventory of important activities, recommendations, techniques, and concerns that are commonly involved in the migration process in one place. This enables academia and practitioners in the cloud computing community to get an overarching view of the cloud migration process. Furthermore, the survey identifies a number challenges that have not been yet addressed by existing approaches, developing opportunities for further research endeavors.

**1 INTRODUCTION**

Many enterprise software applications that support IT services are characterised by the need for high computing capability, scalability, and resource consumption (Buyya et al., 2009, Armbrust et al., 2010). In recent years, cloud computing initiatives have received significant attention towards addressing these requirements through offering services in various forms such as SaaS (software as a service), PaaS (platform as a service), and IaaS (infrastructure as a service) which are universally accessible, acquirable, and releasable in a dynamic fashion, and payable on the basis of service usage amount. Given these advantages, many IT-based organisations have been interested in moving their legacy assets to cloud environments. It is estimated that the global cloud computing market will grow from $40.7 billion in 2011 to $241 billion in 2020 (Ried S, 2011). So far, a significant collection of research has been focusing on this topic by both academia and practitioners ranging from pure technical-centric solutions related to using of cloud services, to research around the social and non-technical impact of the cloud as a new emerging paradigm. However, studies that focus on designing approaches offering a process model (methodology) for the cloud migration have not yet received much attention. Several studies such as (Mohagheghi et al., 2010, Chauhan and Babar, 2012, Jamshidi et al., 2013) suggest that a well-defined process model for supporting migration (or development) and maintaining working legacy applications to the cloud is a key concern. The need for a systematic approach is particularly important when organisations are heavily dependent on legacy applications that have been in operation and stored critical data over the years. Moving to the cloud raises many concerns such as security, interoperability, and vendor lock-in.

Legacy applications often predate the cloud computing and thus have been developed without taking into account the characteristics of cloud environments. The complexity of migration is exacerbated by the fact that some legacy applications may have been developed without taking into account the unique requirements attributed to cloud environments such as elasticity, multi-tenancy, interoperability, and refactoring. Such requirements raise new challenges to the migration of applications to the cloud and hence needs improving conventional software development methodologies to address these specific requirements. Various projects and studies in cloud computing community define migration approaches in order to enable legacy applications to take benefit from cloud services (e.g. reducing maintenance costs, economies of scale, and pay-as-you-go). Although trivial migration projects may be manageable in an ad-hoc manner, a methodological approach becomes important when there is a plan to move large-scale and complex legacy applications that support core business processes of an organisation. A well-structured methodology can aid developers to carry out an effective and safe application migration, instead of struggling to understand "what" and "how" to carry out such transition in an ad-hoc manner which may latter result in poor and erroneous migration and maintenance overhead. A methodological approach can be acclaimed as promising mean for tackling the cloud migration complexities and move from an ad-hoc cloud migration to a structured and step by step quality methodology. In this spirit, Laszewski and Nauduri, who are designer of a methodology for moving Oracle legacy applications to the cloud mention: *Like any software development project, migration projects require careful planning and good methodology to ensure successful execution* (Laszewski and Nauduri, 2011). A similar recommendation can be found in the final report of REMICS project, which is a three years research project supported by the European Commission and focuses on a methodological support for moving of legacy applications to cloud platforms (Benguria et al., 2013). The above report mentions that *in the beginning it [legacy migration] was motivated by the lack of documentation, but in the last years it has been motivated by adaptation to new technologies. Each new technology has required new and renewed approaches and technologies to address the migration process in a more effective way*.

As will be elaborated in Section 2.3, there are several surveys on the cloud migration each focuses on different aspects of cloud migration such as interoperability, techniques and tools for migration, and cloud architecture design. Although these surveys provide a partial understanding of certain aspects of legacy to cloud migration, they do not provide a complete picture of how the cloud migration is to be carried out and organised from the process model perspective. There is not yet a rigorous analysis of the extant material on this aspect of the cloud computing. For this reason and regarding the fact that the interest for legacy application migration to the cloud grows, there is a need to contribute a survey that distills existing cloud migration approaches by identifying their common characteristics and varying motives, concomitant activities, and empirical findings. This survey will differ from existing related surveys (Section 2.3) by focusing on the process aspect of the cloud migration to understand what essential activities and concerns are involved during the legacy to cloud migration. By comprehensively reviewing existing cloud migration approaches, we thus position this survey as the newest reference point for the cloud computing research and practice. Accordingly, the current study will attempt to answer the following research questions:

- **RQ1.** What are the existing approaches proposing a migration model for moving legacy applications to cloud environments in the literature?
- **RQ2.** What is the current state of these approaches w.r.t. the proposed evaluation framework introduced in Section 3?
    **RQ2.1.** what generic criteria, as typically expected for a software development methodology, are supported by these approaches?
    **RQ2.2.** what cloud-specific criteria are supported by these approaches?

RQ1 is motivated by the need to describe the state-of-the-art of cloud migration approaches. This gives readers an overall understanding of the approaches' idea, their core objectives, and a concise description of them. RQ2 was formulated to characterise as well as highlight the focus of approaches with respect to two dimensions. (See Section 3). More specifically, RQ2 aims to answer two sub-research questions RQ2.1 and RQ2.1.

RQ2.1 assesses generic criteria that any process model would need to address regardless of its application genre. RQ2.2 is related to the evaluation of cloud-specific aspects of migration approaches. This decomposition is a first step in the synthesis of the evaluation framework which we will later use to identify and highlight a rich collection of key activities and recommendations that existing approaches include. In summary, the contributions of this paper are the following:

- To provide a deep understanding of the current state of migration approaches proposed in the literature, understand insightful activities and recommendations to be learned,
- To help both researchers and practitioners in the cloud community if they want to capture key facets of existing approaches and select or discard one or collection of them that may suite their needs for a particular migration exercise, and
- To give a broad view of research challenges, specifically concerned with process models for the legacy to cloud migration that need to be investigated by researchers. Hence, a gateway to new research opportunities can be opened.

This paper is structured as follows: In Section 2, we give a general review of terms related to cloud migration, key challenges that need to be addressed in a migration process, and the related work to this paper. Section 3 describes proposed evaluation framework designed for the purpose of this paper. Section 4 presents the research methodology. Section 4 reports the findings of the review after identifying the existing significant approaches. Section 5 discusses the remaining challenges and promising directions for future research. Section 6 presents the limitations of this survey. Finally, section 7 concludes this paper.

## 2 BACKGROUND AND RELATED WORK

As with all new areas of study, an etymological analysis is instructive. This is first undertaken in this section to give some clarity as to what a cloud migration methodology might mean in the context of cloud computing. This section then identifies technical and organisational concerns of such a methodology and provides a review of surveys related efforts.

### 2.1 ETYMOLOGY

—*Cloud migration methodology*. In software engineering (SE) a software development methodology can be defined as a *systematic way of doing things in a particular discipline* (Gonzalez-Perez and Henderson-Sellers, 2008). Another definition can be borrowed from Avison and Fitzgerald: *a recommended collection of phases, procedures, rules, techniques, tools, documentation, management and training used to develop a system* (Avison and Fitzgerald, 2003). A methodology organises the coordination of development team members and integration project activities. It defines when a certain activity, which contains sequence and input/output artefacts, should be carried out.

> Migration of legacy applications to the cloud signifies that the organisation has already in place existing software applications earmarked to take advantages of cloud services. A common understanding of the term *cloud migration methodology*, as offered by (Chauhan and Babar, 2012), is the re-engineering process of legacy applications for becoming cloud-enabled. That is, migration to cloud is a kind of software reengineering where the target application will be able to interact or become integrated with cloud services. Another definition, offered by Andrikopoulos, views the cloud migration process as a set of architectural adaptations required to ensure a legacy application becoming cloud-compliant (Andrikopoulos et al., 2013). Similarly, Kwon et al. pose the term *cloud refactoring* in which code transformation mechanisms are used to integrate legacy applications and cloud services (Kwon and Tilevich, 2014). Another yet broader and workable definition, which covers both technical and non-technical aspects of the cloud migration is suggested by (Pahl et al., 2013) as: *A cloud migration process is a set of migration activities carried to support an end-to-end cloud migration. Cloud migration processes define a comprehensive perspective, capturing business and technical concerns. Stakeholders with different backgrounds are involved.*

One can envisage a cloud migration methodology as an extended traditional software development methodology to enhance its capability to support cloud computing.

—*Legacy Application.* This paper focuses on approaches addressing the migration of legacy applications to cloud environments. As such, we also analyse the term *legacy*. In software engineering literature many definitions can be found for the term legacy applications. One of the earliest definitions is the following: *large software systems that we don't know how to cope with but there are vital to our organization* (Bennett, 1995). Similarly, Stonbraker mentions that a *legacy application is any system that significantly resists modification and evolution* (Brodie and Stonebraker, 1995). Sneed states they are *information systems that have been in use for years* (Sneed, 2006). Others emphasise technological aspects. E.g1, Stone distinguishes legacies as those that are not Internet-dependent (Stone, 2001). E.g2, Dedeke defines it as *an aggregate package of software and hardware solutions whose languages, standards, codes, and technologies belong to a prior generation or era of innovation* (Dedeke, 2012).

A common finding in all above definitions is a dialectic tacit dependence and acknowledgment of the worthiness of the legacy applications. Holland explicitly mentions that *legacy applications encapsulate the existing business* *processes, organization structure, culture, and information technology* (Holland and Light, 1999). Likewise, legacies have been characterised *as massive, long-term business investment, and crucial to an organization* (Bisbal et al., 1999). Legacies are one of the major components of organisations, represent business services and repository of knowledge of organisations, and they can provide a significant competitive advantage with a positive return and contributing to the organisation revenue and growth (Bennett, 1995, Sneed, 1995, Erlikh, 2000).

From the technical point of view, the term legacy application are often collocated with very old generation of technologies, standards, protocols and programming languages such as FORTRAN, COBOL, or C languages, and old indexed database and file systems. They are often associated with old mainframe applications, with hardware support and operational costs that are responsible for enormous transaction processing and supporting thousands of users and concurrently accessing numerous resources. Nevertheless, modern client-server software applications, which have been developed using the latest tools and technologies available in the marketplace such as .Net Framework and J2EE, but currently do not satisfy new business requirements are considered as legacy (Khadka et al., 2013, Sneed, 2006). A common architecture style of enterprise applications is 3-tiered, i.e. a user interface tier, a business logic tier, and data tier. Each tier can have multiple components which can be deployed in different servers and collaborate together.

—*Different types of legacy migration to the cloud*. Taking into account the major cloud service delivery models, i.e. IaaS (infrastructure as a service), PaaS (platform as a service), SaaS (software as a service), one can view there are several possibilities that legacy applications can utilise cloud services. In this survey these are called variant types of the legacy to cloud migration. From this angle, a cloud migration methodology can then be viewed as a systematic process model to perform one or more migration type(s). Inspired by the classification introduced in (Andrikopoulos et al., 2013), Table I presents definitions and examples of some variants. Each migration type may raise different concerns. For example, if a migration type II is intended, where a legacy application is re-engineered to SaaS, then multi-tenancy aspects such as application customisability and resource provisioning are needed to be properly addressed by the application owner. However, in the case of encapsulating an application into a virtual machine and deploying it in the cloud (migration type V), enabling the feature multi-tenancy might be of less or not concern. As another example, the migration of a legacy relational database to a NoSQL cloud database service (type IV) may raise incompatibilities issues between functionalities of relational database tier and equivalent ones offered by NoSQL cloud database. Hence, various approaches may be designed to define process model for a particular migration type while ignoring the others.

Table I. Migration types and classification of approaches based on migration type

| Migration | Symbol | Migration Type Definition |
|---|---|---|
| Type I | 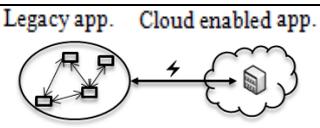 | Deploying the business logic tier of legacy application (e.g. WS-BPEL), which offer independent and reusable functionalities, in the cloud infrastructure by applying the delivery model IaaS. In this migration type, the data tier is remained in local organization network. Deploying an image-processing component of an application in E2C, is an example of this migration type. |
| Type II | 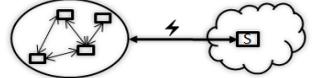 | Replacing some components or whole legacy application stack with an available and fully tested cloud service, by applying the service delivery model SaaS. The Salesforce CRM application is a typical example of SaaS, which can be integrated with other applications via its interfaces. |

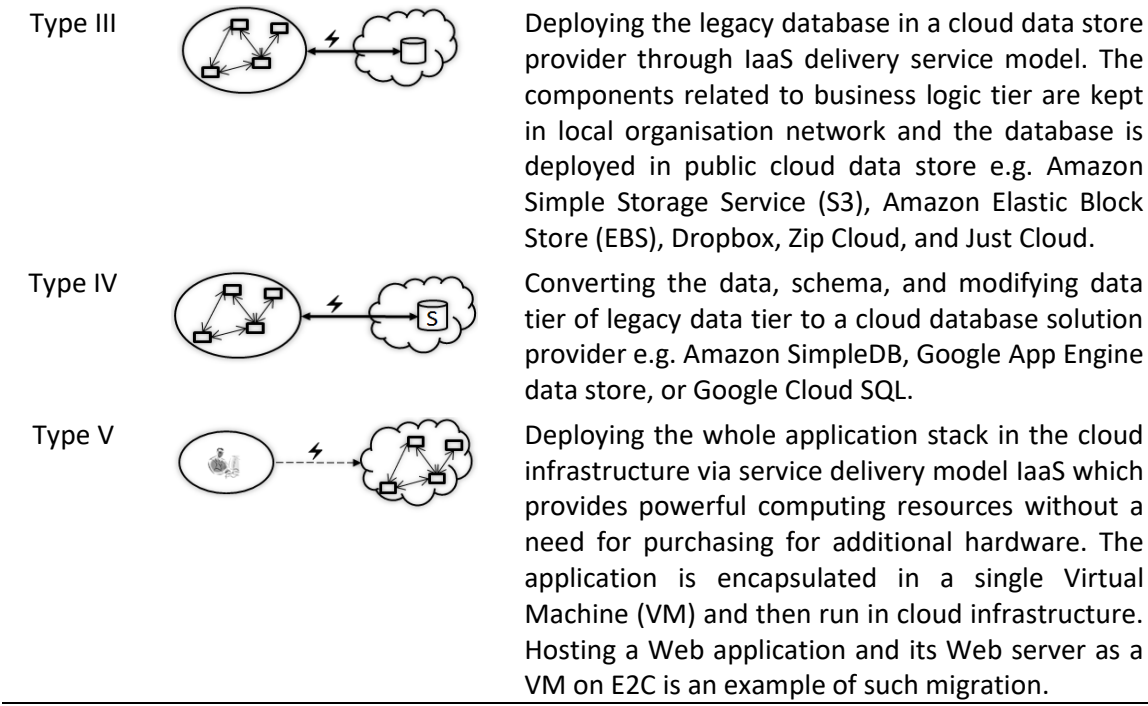

| | | |
|---|---|---|
| Type III | | Deploying the legacy database in a cloud data store provider through IaaS delivery service model. The components related to business logic tier are kept in local organisation network and the database is deployed in public cloud data store e.g. Amazon Simple Storage Service (S3), Amazon Elastic Block Store (EBS), Dropbox, Zip Cloud, and Just Cloud. |
| Type IV | | Converting the data, schema, and modifying data tier of legacy data tier to a cloud database solution provider e.g. Amazon SimpleDB, Google App Engine data store, or Google Cloud SQL. |
| Type V | | Deploying the whole application stack in the cloud infrastructure via service delivery model IaaS which provides powerful computing resources without a need for purchasing for additional hardware. The application is encapsulated in a single Virtual Machine (VM) and then run in cloud infrastructure. Hosting a Web application and its Web server as a VM on E2C is an example of such migration. |

## 2.2 KEY CONCERNS IN APPLICATION MIGRATION TO CLOUD ENVIRONMENTS

Moving applications to the cloud is similar to conventional legacy application re-engineering. However, cloud applications should also satisfy specific cloud environmental concerns. Drawing on the general literature on cloud computing (Fox et al., 2009, Brebner, 2012, Rimal et al., 2009, Guo et al., 2007, Nathuji et al., 2010, Toosi et al., 2014, Dalheimer and Pfreundt, 2009, Ristenpart et al., 2009), we identified six cloud intrinsic key concerns as follow: (i) resource elasticity, (ii) multi-tenancy, (iii) interoperability and migration over multiple-clouds, (iv) application licensing, (v) dynamicity and unpredictability, and (vi) legal issues. These concerns trigger considerations that an application owner should consider them in the migration process, though they might have been already automatically supported by cloud providers. The remainder of this section delineates these six concerns.

**(i) Resource elasticity.** Cloud environment can be viewed as an infinite pool of resources such as CPU, memory, storage, and network bandwidth in a way that resources can be acquired and released by applications based on demand (Fox et al., 2009, Brebner, 2012). Nevertheless, running an application on the cloud does not provide elasticity per se, rather applications need to optimise resource usage in the case of fluctuation in their workload in order to reduce infrastructure cost. Many legacy applications might not have been implemented with a support of dynamic scaling up/down of resources. They assume that elasticity is supported by providing more powerful physical servers. Inevitably, the architecture refactoring of these applications will not be easy and force many modifications in the application tiers. Addressing resource elasticity is concerned in migration types I, II, and V.

**(ii) Multi-tenancy.** In cloud environments, each service consumer is called a tenant (Rimal et al., 2009, Guo et al., 2007). Multi-tenancy is an ability to use the same instance of a resource at the same time by different tenants. On the side of cloud service provider, this maximises the resource utilisation and profit since only one application instance is required to deploy in the cloud. On the cloud consumer side, each consumer feels that he/she is the only users of the application. Tenants can customise application components, such as user interface appearance, business rules and sequence of workflow execution, and last but not least the application code.

However, migration from single-tenant to multi-tenant raises several issues, specifically for migration type II. As tenants are dedicated to the same instance of an application, there is often a concern that tenant's QoS is negatively affected by other tenants. For example, the performance of a tenant that uses one core of a multicore processor may significantly be reduced when another tenant runs an adjacent core and performs a massive workload (Nathuji et al., 2010). This issue can also be re-interpreted from a security perspective where all the tenants may use the shared resources and a malicious tenant with a criminal mind can damage resources and pose a serious threat to all the tenants. Tenant isolation for QoS satisfaction (e.g. performance, security, availability and customizability) should be carefully addressed in a cloud application.

**(iii) Interoperability and migration over multiple-clouds**. The cloud environment is proliferated with numerous services which bring a wide range of possible building blocks to develop a fully-fledged cloud application. Interoperability becomes an issue if an application is built via a composition of cloud services that are developed by different providers, for example, a composition of PaaS and SaaS, whilst each provider has its own particular APIs and interfaces for access to services, causing APIs mismatch (Toosi et al., 2014). These issues face developers to heterogeneities across the application tiers, which imply a certain level of development effort, specifically in migration types I, II, III, IV, and V. As advancements in the cloud computing is still on on-going track and there is not a common standard for development cloud services, application portability is a challenge when its components are to move from a provider to another provider, but there is an incompatibility between underlying technologies of these providers (e.g. APIs). Lack of a proper interoperability can also limit flexibility to use various cloud services where each service provider proposes proprietary interfaces to access services.

(**iv) Application licensing**. Taking advantage of elasticity for applications can raise application licensing issues. For example, assume an organisation that has contracted for K number of application licenses and pays a fixed annual fee. Once encapsulated into a virtual machine and run in the cloud, multiple instances of it are created by a server based on the workload. In such, the restriction on K instances is violated. It seems that traditional licensing model for commercial applications is not workable in cloud-based software. In some cases, a dynamic licensing mechanism is implemented in the application by its owner (Dalheimer and Pfreundt, 2009). However, dynamic licensing is not often addressable in the application and, instead a negotiation between the application owner and provider is required, in particular for migration types I, II, and V.

**(v) An unpredictable environment**. Unpredictability is another big challenge when moving applications to the cloud. As an example, the sudden crash of Amazon EC2 cloud in 2011 caused the website of several high-profile companies down for hours (Blodget, 2011). Data loss was minor though it could be very harmful. Developers should empower the application with proper countermeasures to deal with such behaviors though in some cases unpredictability can be out of control of either application owner or cloud provider.

**(vi) Legal issues**. Beyond the technical aspects, it is a usual concern at the government or organisation level about where application components are hosted and processed (Ristenpart et al., 2009). In the cloud, there is no longer an assumption about the location of components. They may move between different cloud infrastructures based on the server workload and load balancing mechanism defined by the cloud provider/consumer and network traffic. Hence, moving applications to the cloud become problematic from the security perspective. To protect application components for confidentiality, developers need to provide mechanisms in the application to ensure the security of sensitive data within legal boundaries, specifically those components may be relocated between servers in different geographical areas, specifically for migration types I, II, III, IV, and V.

## 2.3 RELATED SURVEYS

While there are several published surveys on different aspects of cloud computing, to the best of our knowledge, there is not a survey devoted to review the literature on existing approaches, resulting in a process model for moving legacy applications to the cloud. Perhaps, the closest studies to this paper are those related to migrating legacies to SOA because SOA and cloud computing share similar characteristics such as using services as basic blocks to build reliable and secure applications (Yi and Blake, 2010).

Razavian and Lago report a systematic literature review of SOA migration approaches (Razavian and Lago, 2015). Their key goal of the survey is to identify commonalities and differences between 75 identified approaches and propose a reference model of typical activities that are carried out for the legacy to SOA migration. From the knowledge management point of view, this reference model includes typical knowledge that shapes a process of evolving legacy applications to service-based applications. Khadka et. al. provide a historic review of methodologies for the legacy to SOA migration (Khadka et al., 2013). The objectives of this review are (i) to reach a broad understanding of existing process models for legacy evolution to SOA (ii) identify available techniques to perform migration activities, and (iii) identify the existing issues and possible directions for future research. Through evaluating 121 primary studies using an evaluation framework, inspired from three traditional reengineering methodologies namely Butterfly (Bisbal et al., 1997), Renaissance (Warren and Ransom, 2002), and Architecture-Driven Modernization (ADM) (Khusidman and Ulrich, 2007), the authors conclude that there is still a lack of adequate automation level and techniques for determining the decomposability of legacy applications, investigating organisational perspective of migration, and postmortem reports on after-migration experience. In another attempt, Lane et. al. present a survey of process models to develop service-based applications with a skew towards dynamic adaptation (Lane and Richardson, 2011). Subsequently, they developed a meta-model giving an overarching view of development processes of 75 identified methodologies. On the basis of evaluation results using this meta-model, they found that increasing the automation of development process using model-driven development techniques is the most common theme in the reviewed methodologies. They also argued that existing methodologies suffer from a lack of real empirical validation. Even though the above-mentioned surveys are helpful, they are silent to address the cloud-centric challenges stated in Section 2.2.

(Jamshidi et al., 2013) identified: (i) the main drivers which motivate organizations to move their legacy applications to the cloud, (ii) different types of migration activities might be performed, (iii) techniques and tools, and (iv) existing gaps in the literature. Twenty one studies on cloud migration were evaluated against a characterisation framework including contribution type, evaluation method, means of migration, migration type, migration tasks, intents of the migration, migration tool support, and constraints. In another review, the REMICS consortium presents the state of the art with respect to modernisation methodologies and tools (SINTEF, 2011) that support the automatic transformation of legacy application components to cloud environments in a model-driven fashion. However, none of the preceding surveys place particular focus on approaches for moving legacy applications to the cloud. A number of other review papers were also identified but they fall outside the scope of this survey. For example, (Girish and Guruprasad, 2014) focuses on six data frameworks for cloud migration. Furthermore, (Medina et al., 2014) reviews research related to live migration mechanisms of virtual machines which move from one host to another over cloud network. Other studies such as (Singh and Chana, 2012) and (Taher et al., 2012) report general challenges of cloud-based application development in terms of cloud-based architecture, component-based development and reusability, quality, design, and security.

The survey provided in the current study is different from the existing reviews in three salient aspects. Firstly, this survey limits its focus on all extant approaches proposing a (complete or partial) migration process model or framework for the cloud migration, and hence is more specific than the above-mentioned surveys. None of the reviewed surveys (see Table II) provides an in-depth discussion on the features and migration activities proposed in the existing approaches as well as useful experience of applying these approaches in practice. Secondly, this survey provides an in-depth analysis of existing approaches through an evaluation framework, which encompasses 28 criteria classified into two dimensions i.e. generic and cloud-specific ones. The proposed framework has been derived through an extensive literature review and validated through a Web-based questionnaire survey of 104 experts from academia and experts in the field of cloud computing. Since all related surveys fail to consider the important evaluation criteria that the proposed framework includes, the evaluation framework can be considered as an important contribution of the current study. The characterisation framework of (Jamshidi et al., 2013) does not include any generic criteria as offered by our proposed evaluation framework. For the cloud-specific dimension, although 11 criteria have been referred by their framework, there is no elaboration on assessment of approaches. Thirdly, given our different focus, none of the related work covers the papers that this paper reviews. We found that only 5 out of our 43 reviewed papers were covered by (Jamshidi et al., 2013). Finally, this survey considers different and recently published approaches that are not covered in the other surveys. With respect to this, this survey can be viewed as complementary one to the above-mentioned surveys through investigating different and recent approaches.

Table II. Related surveys on SOA/Cloud migration

| Authors and title | Study type | Published channel | Aim | Publication year | Total reviewed | Time period of included studies |
|---|---|---|---|---|---|---|
| M. Razavian, P. Lago, "*A systematic literature review on SOA migration*" | Systematic literature review | Journal of software: evolution and process | To review and analyse existing SOA literature in order to identify commonalities and difference between designed migration approaches. It designs a new conceptual model, including a classification of activities carried out for the legacy to SOA migration. | 2015 | 75 | 2003-onwards |
| R.Khadka, A. Saeidi, "*Legacy to SOA Evolution: A Systematic Literature Review*" | Systematic literature review | Book chapter | To provide a broad understanding of process model required for legacy evolution to SOA, available techniques to conduct migration activities, existing issues, and possible directions for future research. | 2012 | 121 | 2000- August 2011 |
| S.Lanea,I.Richardsona, "*Process Models for Service Based Applications: A Systematic Literature Review*" | Systematic literature review | Information and Software Technology | To assess existing process models for service-oriented application development on the basis of a generic process metamodel and identify categorised activities within those process models focus area of existing methodologies. | 2011 | 57 | October 2009 |
| P.Jamshidi, A.Ahmad, "*Cloud Migration Research: A Systematic Review*" | Systematic literature review | IEEE Transactions on Cloud Computing | To identify the main drivers which motivate organisations to migrate their legacy applications to the cloud, existing approaches and different types of migration activities, techniques, and tools for migration. | 2013 | 21 | 2005 - 2013 |
| REMICS Consortium "*State of the art on modernization methodologies, methods and tools*" | Review | REMICS (Public deliverable) | To review traditional modernisation methodologies and tools and argue potential requirements for designing cloud migration methodologies. | 2011 | NA | NA |
| Girish, "*Survey on Service Migration to Cloud Architecture*" | Review | Journal of Computer Science & Engineering Technology | To identify commonalties and difference of some existing cloud migration methodologies. | 2014 | 22 | NA |
| V. Medina, J. Garc, "*A survey of migration mechanisms of virtual machines*" | Review | ACM Computing Surveys | To identify live migration mechanisms of virtual machines which move from one cloud to another over the cloud network. | 2014 | 14 | NA |
| S.Sukhpal, C. Inderveer, "*Cloud Based Development Issues: A Methodical Analysis*" | Review | Journal of Cloud Computing and Services Science (IJ-CLOSER) | To identify general issues in development cloud-based applications in terms of cloud-based architecture, component-based development and reusability, quality, design, and security. | 2013 | 89 | NA |
| Y. Taher, D. Nguyen, "*On Engineering Cloud Applications - State of the Art, Shortcomings Analysis, and Approach*" | Review | Scalable Computing: Practice and Experience | To present a review of the standardisation, methodology, software, and products that support development of service-based applications in the cloud and state existing gaps in the literature. | 2012 | NA | NA |

# 3 EVALUATION FRAMEWORK

We propose an evaluation framework leaning heavily towards assessing software development methodologies, allowing us to classify and characterise approaches applicable to cloud migration and answer to our research questions. The following meta-criteria, suggested by Karam et. al. (Karam and Casselman, 1993), were used to define a fair evaluation framework that is: (i) sufficiently general to all methodologies, (ii) precise enough to characterise the similarities and differences of methodologies, (iii) and comprehensive enough to cover all important requirements of methodologies. A two-dimensional set of criteria have been developed: generic criteria and cloud-specific ones as shown in Fig. 1.

For the generic criteria dimension, we reviewed and synthesised various existing frameworks that define criteria attuned to evaluate software development methodologies in software engineering. These frameworks were (Karam and Casselman, 1993), (Wood et al., 1988), (Ramsin and Paige, 2008), (Sturm and Shehory, 2004), and (Tran and Low, 2005). After analysing these sources, removing redundancy and overlapping between criteria, 11 distinct criteria were derived for the purpose of this study including (1) Process clarity, (2) Procedure and supportive techniques, (3) Tailorability, (4) Development roles, (5) Modelling language, (6) Traceability, (7) Work-products, (8) Formality, (9) Scalability, (10) Tool Support, and (11) Domain applicability. A detailed description of these criteria is presented in Appendix F. Section 5.3 motivates and elaborates each criterion by providing a detailed explanation for each, along with an evaluation result against existing approaches.

For the other dimension, the framework was expanded with cloud-specific criteria that were deemed important and relevant to legacy application migration to the cloud. This was initially inspired by the study introduced in (S. Strauch, 2014) and (La and Kim, 2009) that proposed a small set functional and non-functional properties that should be addressed by an ideal cloud migration methodology. These studies, however, were not complete and well-articulated; nor enough attention was paid to domain-independence, validation, and generality. Hence, we strived to identify a coherent set of analysis criteria for inclusion in the evaluation framework and accordingly to assess cloud migration approaches. The derivation of the criteria was mainly influenced by various sources in the cloud migration literature, specifically those proposed in (Strauch et al., 2014) and (Jamshidi et al., 2013), and the reviewed approaches in this survey.

A criterion is included in the proposed framework if it had addressed at least one of the concerns stated in Section 2.2 and also sufficiently generic to cover a variety of migration scenarios regardless of a particular cloud platform. This resulted in defining 17 cloud-specific criteria including (1) Analysing Context, (2) Understanding Legacy Application, (3) Analysing Migration Requirements, (4) Planning Migration, (5) Cloud Service/Platform Selection, (6) Training, Re-Architecting Legacy Application (including (7) Incompatibility Resolution, (8) Enabling Multi-Tenancy, (9) Enabling Elasticity, (10) Cloud Architecture Model Definition, (11) Applying Architecture Design Principles), (12) Training, (13) Test and Continuous Integration, (14) Environment Configuration, (15) Continuous Monitoring, (16) Migration Type, (17) Unit of Migration. These criteria helped the study to contrast and compare cloud-centric aspects of existing approaches. Once these criteria were established, their importance and relevance to the cloud migration were assessed through a Web-based survey of 104 experts from academia and practitioners in the cloud computing field. This gave confidence that important criteria are covered by the framework. For further detail, we refer to the report (Fahmideh, 2015) which provides a detailed analysis of the criteria and suggested comments about the criteria by the survey respondents. The above report also includes the statistical analysis of importance of each criterion and the qualitative comments of experts to justify the importance of each criterion. The criteria are defined in Appendix H and illustrated in Section 5.4.

Since the criteria were aimed to be measurable, evaluation questions were defined for each criterion, with a four possible kinds of answer for each as suggested by (Kitchenham et al., 1997): Narrative, YES/NO, Scale, or Multiple choices. For open questions a narrative answer was used; a YES/NO answer was given where a criterion was met or not; an answer that is described by scale point refers to a degree of criterion support provided by an approach. Scale points were used during evaluation to assign a value for the level of support of a criterion. Here, the framework defines three levels of scale as fully-supported (●), partially-supported (◓), and not-supported (○). Finally, multiple choices were defined when there was a possibility to choose one out of a number of answers (for example, the criteria migration type).

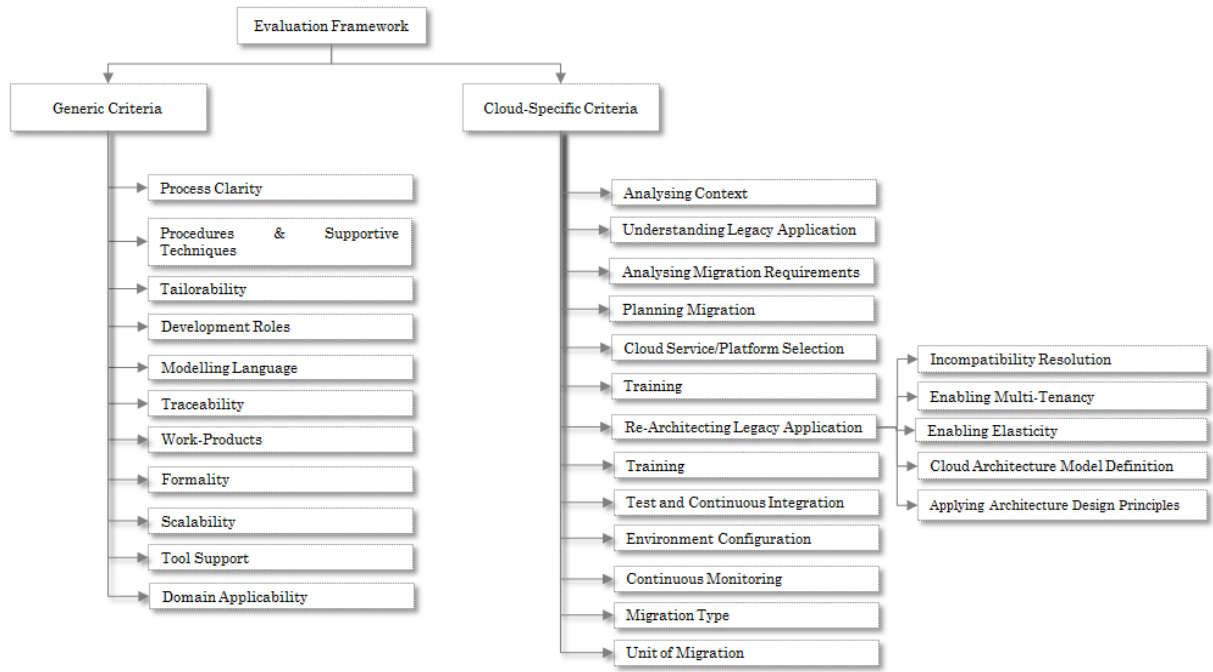

Fig. 1. Proposed Evaluation Framework for assessing various approaches (See Appendices 'F' and 'H' for the criteria definitions)

**4 SURVEY**

The systematic review procedure proposed by (Kitchenham et al., 2009) was selected as a survey approach. It is a well-recognized procedure to identify, analyse, and interpret all related studies with respect to a topic of interest and includes three phases 'Planning', 'Conducting Review', and 'Documenting Results'. In the planning phase, a protocol for the review is established which includes the defining of search strings, study sources, and study selection and inclusion criteria. These definitions are used to conduct the next phase, which is the review phase. During the review phase, various studies are identified, analysed against the inclusion and exclusion criteria, and necessary data items are extracted from them. In the last phase, the results of the review are documented. Appendix A describes steps in each phase. Consequently, 43 papers were identified, as the output of this step, for the review after applying the inclusion and exclusion criteria. For more detail See Appendix B.

**5 RESULTS**

Through analysing 43 identified approaches, we answered three research questions stated in Section 1. Section 5.1 presents (i) overall demographic information about the studies including the year of publication, publication channels, and authors' nationality and (ii) the result of the quality assessment of the papers. Sections 5.3 and 5.4 describe the analysis of the studies against each dimension of the proposed taxonomy.

## 5.1 OVERVIEW OF APPROACHES

**Quality Assessment.** While any new research that contributes to the literature is regarded as important, its findings might not be reliable if it suffers from a lack of appropriate research quality (Kitchenham et al., 2002). Nine criteria provide a measure of the extent to which a research could contribute to the literature. Table I.1 (Appendix I) shows how each study is assessed by the criteria. Given the questions in Appendix C, the grading of each criteria was based on three scales: 'fully supported', 'partially supported', and 'not-supported'. Fig. 2 depicts the distribution of the identified approaches in terms of their support for the criteria.

According to Table I.1 the majority of the approaches have clearly stated a research aim. As many as 25 out of the 43 approaches have provided contextual information about the environment in which the research had been conducted. Seven approaches stated a general description of the research context whilst 11 ones did not state any information. As far as the criterion 'research design' is concerned, only 6 studies (14%) [S4], [S15], [S17], [S20], [S31] and [S36]) provide a clear description of the research design of the study. As many as 12 of the 25 studies did not sufficiently describe the research design. Additionally, an overall view of the scores in Table I.1 (Appendix I) reveals that a large portion of the studies have not addressed the criteria 'Data Collection', 'Data Analysis', and 'Reflexivity'. More exactly, as many 30 and 31 studies ignored to report how data had been collected and analysed for the proposed approach. It can be seen that 39 studies (90%) did not report how researchers have been involved with validation environment. From this observation, one may conclude that the research methodology has been viewed as a subsidiary by the designers of approaches. Thus, future approaches deserve attention to methodical research. We discuss this issue in Section 6. Back to this table it can be seen that 17 approaches explicitly stated the contributions of their research. Finally, only two studies (5%) [S15], and [S31] achieved a full score on the quality assessment, and two studies [S1] and [S7] received the lowest score in this review.

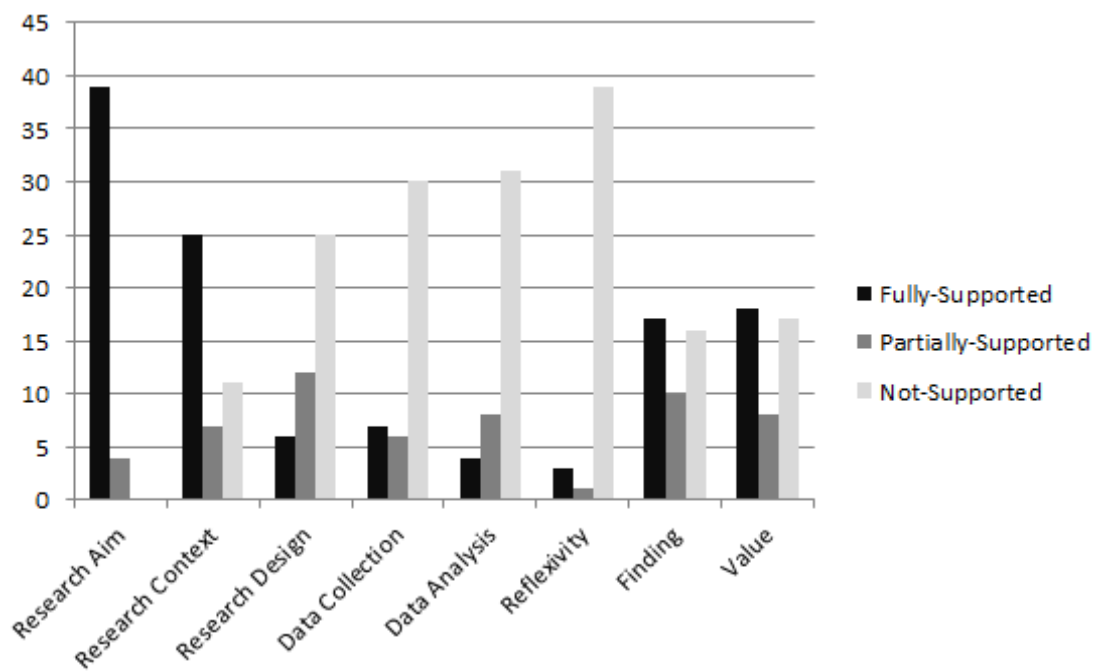

Fig. 2. Quality scores for the identified studies

For the criterion 'Validation Type', the studies were classified according to the applied validation type as shown in Table VIII.1. The majority of approaches applied the case study technique to evaluate their work (21 studies), followed by techniques such as interview (4), industrial experience (4), example (3), focus group (2), simulation (2), theoretical evaluation (1), and questionnaire (1).

Four of the studies (i.e. [S1], [S15], [S18], and [S27]) used a combination of validation types. For example, [S1] reported a case study in addition to a questionnaire. Of 43 approaches, 10 (23%) did not report any information about validation of suggested methodology.

Table VIII. Validation type

| Study | Validation type | Description | Number |
|---|---|---|---|
| [S1], [S5], [S6], [S7], [S8], [S9], [S10], [S12], [S14], [S15], [S16], [S17], [S20], [S22], [S23], [S25], [S27], [S29], [S30], [S37], [S39] | Case Study | Case study has been used to evaluate the approach. | 21 |
| [S1] | Questionnaire | Experts were asked to answer questions about the suitability and applicability of the designed migration approach. | 1 |
| [S4], [S15], [S18], [S34] | (Semi-Structured) Interview | Experts were interviewed about the quality of the designed approach. | 4 |
| [S18], [S34] | Focus Group | A group of experts were asked about their opinions towards the approach in an interactive group setting. | 2 |
| [S2], [S3], [S32] | Example | Application of the approach has been demonstrated through an illustrative example. | 3 |
| [S11], [S36] | Simulation | A mathematic simulation has been used to assess the approach correctness. | 2 |
| [S7], [S13] | Theoretical Evaluation | The approach has been evaluated using a set of high-level criteria. | 1 |
| [S26], [S33], [S35], [S43] | Industrial Experience | The approach has been developed on the basis of gained experience in an industrial experience. | 4 |
| [S19], [S21], [S24], [S27], [S28], [S31], [S38], [S40], [S41], [S42] | Not stated | The approach did not specify any applied validation. | 10 |

**5.2 RQ1 What are existing approaches proposing a migration model for moving legacy applications to cloud environments?**

Given a scanning of the approaches, Appendix D synopsises a description of the approaches, ascendingly sorted based on published year. It does not aim to present a critique on the existing approaches, but instead it abstracts the approaches and gives a broad understanding of their perspectives to a cloud migration process, and facilitates for further investigation, elaboration and improving their process. In this table, the third column presents the theoretical foundations used to design the approaches. According to the third column of this table, the motivation of the existing cloud migration approaches varies between nine different streams as presented in Table XIV and described in the following.

—*Model-Driven Development*. The most applied paradigm is model-driven development with 7 approaches. Its primary goals are portability, interoperability, reusability of applications as well as and increasing development speed. Applied in cloud migration, it meant to transform the legacy application models (e.g. codes and architecture) into platform-independent models, configure them and then generate platform-specific cloud applications using model transformation techniques. As an example, REMICS [S26] is a model-driven methodology with a special emphasis on cloud-enabled legacy applications that can be run on multiple clouds.

—*Software-Product Line*. In general, SPL-based approaches assume a closed world. Two approaches, [S13] and [S15], incorporate the notion of Software Product Line (SPL) to make legacy applications cloud-enabled for various user requirements. In [S13] authors propose a five-phased process model which incorporates software product line techniques such as domain analysis and variability modeling into SaaS migration. The approach highlights commonality and variability in legacy applications and customises them for various cloud environments. Likewise, the approach suggested in Guillén et. al. [S15] uses a combination of software product line and model-driven development. It defines the notion of *cloud artifacts* which are high-level models of target cloud application. Cloud artefacts are developed based on cloud variability models and transformed to multiple cloud platforms.

—*Agile Development*. Incorporating Agile practice such as light process, short release, and continuous testing into cloud migration approaches have also received attention from the cloud community. In the approach proposed by Krasteva and Stavru [S1], authors pose whether legacy modernisation processes can benefit from Agile methodologies. As a suggestion, they define a Scrum-based extension of REMICS methodology.

—*Software Patterns.* Patterns are reusable and good-enough abstract solutions for recurring problems during the software engineering. They can be used for developing applications through an appropriate composition of their instances. Applied in cloud migration, Jamshidi et. Al. [S20] proposed 15 fine-grained patterns that are identified based on empirical evidence from migration projects. The patterns can be viewed as a sequence of adaptation activities that are performed to modernise legacy application components.

—*Ontology-Based Reengineering.* Ontologies express an area of interest in a communicable and formal way. The ontology-based approach offered by Zhou and Yang in [S39] is a re-engineering process thereby building an ontology of enterprise legacy applications and then decomposing it into potential service migrated to the cloud.

—*Supply Chain Lifecycle*. Lindner et. al. [S38] believe that cloud migration involves different interconnected cloud service providers and consumers that form a supply chain model. A service is provided at the start of the supply chain and a consumer at the end uses this service. A proper cloud supply chain model is required to understand requirements of both providers and consumers during this end-to-end migration process. Given that, they define a breaking down all of the activities and steps, which allows an organisation to migrate legacy applications to the cloud whilst reducing the risk to the business.

Table XIV. Classification of approaches based on theoretical foundation

| Foundation | Approach | Number |
|---|---|---|
| No specific foundation | [S2], [S3], [S6], [S7], [S8], [S9], [S10], [S12], [S16], [S17], [S18], [S19], [S21], [S22], [S23], [S24], [S25], [S27], [S28],[S29], [S30], [S33], [S34], [S35], [S36], [S37], [S40] | 28 |
| Optimization | [S11] | 1 |
| IT Capability Maturity Framework | [S4] | 1 |
| Agile Scrum | [S1] | 1 |
| Model-Driven Development | [S14], [S15], [S26], [S41], [S43] | 5 |
| Software Product Line | [S13], [S15] | 2 |
| Software Patterns | [S20], [S42] | 2 |
| Supply Chain Lifecycle | [S38] | 1 |
| Ontology-Based Reengineering | [S39] | 1 |
| Conceptual Model | [S31] | 1 |
| | **Total** | **43** |

## 5.3 RQ2.1 What generic criteria, as typically expected for a software development methodology, are supported by these approaches?

Appendix E presents the extent to which each approach supports five criteria 'process clarity', 'procedure and technique', 'modelling', 'tailorability', and 'tool support'. These criteria were measured using three levels of scale (See Appendix F). The following subsections details the definition of each criterion in the first dimension of the evaluation framework and justifies the situation in which the criterion can be important. The evaluation results against this dimension also reported. For those studies that provided a full support of criteria, we narrowed our analysis to identify any interesting recommendation and reported experience for presenting in this survey.

### 5.3.1. Process Clarity

Approaches differ in clarifying and specifying the detail of phases and activities where some of them offer the most prescriptive ones whilst others, deliberately or not, provide a short description of the activities and leave a space for an arbitrary interpretation of how activities can be carried out. The degree of prescription is a helpful criterion for developers or organisation who are not familiar with cloud computing and requires a step-by-step methodic guidance on what phases and activities should be followed in order to modernise legacy applications. According to Table E.1 (Appendix E), 35 out of 43 approaches were rated as fully-supported as they provide a description of the suggested activities within their approaches. Seven out of 43 approaches provided a general or short description of the phases and activities with no depth, rated as partially-supported. The only study which was rated as not-supported was the conceptual process model suggested by Jamshidi and Paul [S31], listing 20 key activities for a migration process as a result of a literature review on cloud migration but without any definition for them.

### 5.3.2. Procedures & Supportive Techniques

Apart from a need for clear description expecting from an approach, it is desirable that the approach further elaborates its prescribed activities by defining supportive techniques and examples in a way that developers can simply apply the approach. Hence, if developers seek detailed advice on how to perform activities or producing models rather than top-level descriptions, those that properly describe steps to conduct activities take precedence over others. Twenty three of 43 approaches provide examples of how their prescribed activities can be performed. However, 14 approaches provide guidelines for some activities but not for the whole process. It was found that 6 approaches do not offer any advice on performing activities. Some examples demonstrate how approaches offer techniques to conduct activities, though more detail on various supportive techniques is described in Section 5.4. To resolve incompatibilities between legacy applications and cloud services, the suggested approach in [S35], proposes typical patterns such emulators, proxies, and data aggregators in order to migrate data tier of legacy applications to a cloud database solution. As another example, a common concern in designing a new architecture for the legacy application is to define a proper deployment of legacy components over the cloud. With respect to this, the approach proposed by Leymann et al. [S11] refers to this as a *Move-to-Cloud* Problem and formally transform it into a graph partitioning problem and use existing optimisation algorithm (e.g. simulated annealing) to find optimum distribution of legacy application components between different cloud platforms.

### 5.3.3. Tailorability

Like any software development project, a methodology should be adapted for the particular goals and contingencies of a cloud migration project at hand. Hence, the thought of a universal cloud migration methodology can meet all migration project situations is viewed fallacious. While methodology adherence can be beneficial, constructing situation-specific methodologies or tailoring existing ones that meet project characteristics at hand should not be overlooked. As an example in the context of cloud migration, according to (Louridas, 2010) there is a difference between the US and EU for addressing the ultimate to data protection in the cloud. That is, in the US, a cloud provider is responsible for completely data protection whilst in EU, the cloud consumer is final responsible to ensure if the cloud provider satisfies data protection requirements. As this concern can impact migration process, a migration approach, which is used for moving data tier of legacy application to the cloud, has to incorporate security-related activities into its process.

Given that the tailoring is an important part of approach enactment, it deserves to examine if it provides mechanisms or guidelines for its tailoring in a structured manner. The review showed that most of the existing approaches, except for [S9], [S10], and [S26] which partially inform a need for methodology tailoring, are based on a one-fits-all assumption. Proposed methodologies in [S9] and [S10], jointly, refer to tailoring as *methodology extensibility*. REMICS [S26] methodology is stored as a set of method fragments that can be further selected and assembled regarding characteristics of a project at hand. ARTIST methodology [S43] provides a tool which allows the customisation and instantiation of the methodology on the basis of migration requirements. Nevertheless, there is no guidance on the creation of a situation-specific methodology. Following the idea of situational method engineering (Brinkkemper, 1996, Henderson-Sellers and Ralyté, 2010), Jamshidi et. al [S20] pose the idea of an assembly-based approach wherein a methodology is constructed through combining architectural migration patterns. But, there is no further support for methodology construction or tailoring.

### 5.3.4. Development Roles

As new characteristics and activities are introduced to the cloud application development process, approaches are expected to specify what roles, required expertise, and responsibilities are required. This can be helpful for developers who have limited experience in cloud migration and are not quite clear sure about these roles. Furthermore, depending on chosen migration type (Table I), different roles may be required, or existing roles may be tailored. For example, deploying the whole legacy application tiers in the cloud (the migration type V) would need a role who can analyse application workload and data storage growth; whereas in the case of moving only business tier to the cloud (the migration type I), a reverse engineer would be required to discover the logic of legacy code blocks. It deserves approaches define roles and provide guidance on activities and responsibilities associated with those roles in the course of the migration process. An interesting finding in this survey was that, except for the methodology proposed by Chauhan and Babar [S8], the majority of the existing approaches had not defined required roles. In another approach by Pahl [S18], roles are briefly described but have not been linked to the activities. Table XVI shows the description of identified roles in the existing approaches. More discussion on this issue is presented in Section 6.

Table XVI. Identified development roles

| Role | Approach | Responsibility |
|---|---|---|
| Business Analysts | [S8], [S18] | Elaborating requirements, main motivations for cloud migration, objectives to be achieved, and cost analysis. |
| Systems Analysts | [S8] | Analysing technical incompatibilities between legacy application components and cloud services. |
| Project Managers | [S8] | Identification of potential cloud environments. |
| System Architects | [S8], [S18] | Identifying potential cloud hosting solution, analysing technical incompatibilities between legacy application and cloud environments, design potential architecture solutions, evaluation QoS of cloud, and incremental architecture scoping and definition. |
| Developers | [S8] | Implementing or re-factoring the designed solutions for moving the legacy application to target cloud environments. |

### 5.3.5. Modelling Language

An approach may specify a particular notation and semantic rules for expressing the outcome of each migration activity, i.e. models (or work-products). Modelling will increase the automation and productivity of development process. If developers have already been using tools along with a supported modelling language, this criterion can be helpful to select those approaches that can be supported or integrated with existing tools. Modelling examines if an approach uses a particular notion to describe outputs from the activities as well as the degree of support, i.e. partial or whole life cycle. In this regard, 29 approaches did not use or suggest any modeling language to model the output of activities. Twelve approaches apply a modeling language for some activities. Only REMICS [S26] and ARTIST [S43] methodologies were found to fully support this criterion for their entire lifecycles. Besides, approaches vary in applying the type of modelling languages. For example, REMICS [S26] and ARTIST [S43] use UML. On the other hand, some approaches such as [S20], [S29], [S36], and [S40] keep modelling at the level of simple graphical diagrams and texts. Table XI shows the approaches that use a modeling language, whether with a partially-supported or full-supported, along with their aim to use.

Table XI. Approaches incorporating modeling language

| Study | Modelling Language | Aim |
|---|---|---|
| [S1] | UML | Modelling in general |
| [S11] | UML | Representing a deployment model of the application, i.e. re-arrangement and provisioning of the application components in the cloud. |
| [S12] | AST and SoaML | (i) AST (abstract syntax tree) for the source code representation.(ii) SoaML for modelling individual service interfaces, service implementation, and architecture of target application. |
| [S13] | UML | Modelling commonality and variability in the target domain for which SaaS application is migrated. |
| [S15] | Feature model and XSD | (i) Describing cloud platform using feature model. (ii) Describing cloud deployment model using XSD schema diagram. |
| [S17] | Simple Block Diagram | Representing a legacy architecture model and a cloud deployment model of the application. |
| [S20] | Simple Block Diagram | Representing cloud migration patterns to modify legacy application architecture. |
| [S21] | Graph Modeling | Legacy source code representation |
| [S26] | UML | In general |
| [S29] | Simple Block Diagram | Modelling application architecture before and after adding support for multi-tenancy and architecture decoupling. |
| [S36] | Simple Block Diagram | Modelling the deployment of the application in the cloud. |
| [S37] | UML | Using state chart to model how application workload can move from a cloud to another cloud. |
| [S39] | UML and OWL | Representing the legacy application architecture using UML and transforming it to ontology (UML to OWL model transformation). Then partitioning the application ontology into potential service candidates. |
| [S40] | Simple Block Diagram | Modelling legacy application architecture and dependency tree. |

### 5.3.6. Traceability

Traceability refers to the relationships between work-products in the whole process model in the sense that a model becomes a refinement of another model and all models can be traced to high-level requirements. This criterion, which pioneered by object-oriented methodologies, is important in order to understand the linkages of a work-product to its previous and next work-products and also to manage software changes during the lifecycle. Approaches were examined if they specify mechanisms or guidelines to refine abstract models into more detailed models. One interesting observation in this review was that traceability is weakly supported in the existing approaches. Only 5 approaches ([S8], [S11], [S13], [S26], and [S43]) define the chain of work-product changes. For example, in the approach by Leymann, et al. [S11], models of legacy application are gradually enriched by deployment information and labels. Five approaches provide support for traceability between particular activities but not for their whole model: [S12] (Legacy code model → architecture representation → architecture redesign), [S21] (Requirement analysis → Migration plan, Legacy code → Code model → Legacy architecture → Cloud-service architecture model), [S41] (Legacy model → Target architecture → Mapping model → Constraint violation), [S43] (Source code → model understanding). Like many other areas of software engineering development, traceability is a crucial concern and cloud migration approaches need to take it into account.

### 5.3.7. Work-Products (Artefacts)

An integral part of every methodology is to define necessary work-products as the outcome of each activity throughout the lifecycle. An approach that specifies work-products can be adopted easier by an organisation using modeling tools in daily development activities. The approaches were scanned to identify any recommended work-product. It was found that 27 of 43 approaches defined required work-products that should be produced during the migration process. As approaches varied and played English words for naming and words of work-products, the identified work-products were classified based on their similarity, resulting in 9 distinct work-products recommended by approaches. In this regards, the first and third columns of Table XIII, respectively, show the work-product and different naming for that. From this table, the most frequently recommended work-products are *Legacy Application Architecture Model* and *Cloud Architecture Model* as stated in 14 and 12 approaches, respectively. *Migration Plan* is prescribed by 4 approaches.

Table XIII. Methodologies prescribed work-products during migration process

| Work-product | Migration Type | Aim | Referred names in studies | Frequency |
|---|---|---|---|---|
| Requirement Model | All | A set of computational requirements, motivations, and objectives for migration to the cloud. | System Requirements [S8], Recovered System Requirements [S26] | 2 |
| Cloud Provider Profile | All | A profile of potential cloud providers which can satisfy legacy application requirements. The profile characterises the providers in terms scalability (vertical/horizontal), availability, security, inter-operability, and storage capacity. | List of Potential Cloud Environment [S8], A shortlist of suitable suppliers [S4], Selected Cloud Environments [S8], Selected cloud offering model [S43] | 4 |
| Cloud Architecture Model | All | An architecture model which specifies the re-arrangement of legacy application components in the cloud regarding criteria such as network latency, data transfer, regulations, privacy, and geographical locations of components. The model is also indicates which application components are to be migrated to which cloud and which components are kept in the local network. | Cloud Distribution [S11], Cloud Deployment Model [S22], Cloud Deployment Model [S28], Cloud Deployment Architectures, Cloud Component Model [S36], Target Architecture [S41], Cloud Architecture Model [S29], Deployment Diagram [S11], Architecture Diagram [S11], Cloud-service Architecture [S21], Finalized Design Decision And Modified System Architecture [S8], Cloud outsourcing model [S4], Architecture [S18] | 13 |
| Interaction Diagram | All | Representing how application components are interacting in the a-synchronies and loosely coupled cloud environments. | [S36] | 1 |
| Application Variability Models/Templates | Type II | These models capture features/parameters and the variation points of the application components in the sense that application can be configured for different deployment settings and execution in cloud platforms. Variability models facilitate application interoperability between different cloud platforms. | Variability Models [S13], Cloud Variability Models/Feature Model (deployment model, code model) [S15], Variability Model [S29] | 3 |
| Migration Plan | All | A plan to guide how to conduct the migration process regarding migration needs and feasibility analysis. | Migration Plan [S21], Roll-out plan [S38], Migration Plan [S42], Plan [S40] | 4 |
| Cloud Risk Management | All | Remedy actions to handle cloud risks. | A Cloud risk management strategy [S4] | 1 |
| Legacy Architecture Model | All | An architectural description of the legacy application, the dependency between the components, data model, and interfaces. This model helps better understanding of current state of the legacy application and provides an insight of estimation effort to resolve incompatibilities between legacy application and target cloud solution. | Legacy Architecture [S21], Knowledge Model [S26], Application profiling, Application Dependency Mapping, Application Architectures [S27], Application Model [S30], Source Code Ontology, Database Ontology, Enterprise Ontology [S39], Legacy Architecture [S40], Dependency Tree [S40], Taxonomy of legacy artefacts [S43] | 12 |
| License Agreement | All | A signed contract with cloud provider(s) indicating a model for resource consumption cost. | SLA and Pricing document [S4], Pricing Model [S40] | 2 |
| Virtual Machines | Type V | Specifications of running virtual models of application in the cloud. | Implementation Unit (Virtual Machines) [S11] | 1 |
| Certification Model | All | A model which increases the confidence of end-user about the quality of migrated application and its compliance with the cloud. | Certification Model [S43] | 1 |

### 5.3.8. Formality

This criterion examines the extent to which an approach provides mathematical, unambiguous, and precise mechanisms for the modeling language, representing work-products and the relationships between them. The majority of the approaches do not support formalism. The approach by Leymann, et al. [S11] is the only one that supports a formal mechanism for partitioning legacy application components, selecting a subset, and rearranging them into groups that might be provisioned into different clouds. Moreover, Aakash and Ali Babar [S21] provide a set of mathematical operators for transforming legacy code and architecture models to cloud-enabled ones. But the definitions of the operators still required to be completed. Obviously, if developers seeking for an approach that supports mathematical modeling and reasoning, the current literature is silent.

### 5.3.9. Theoretical Foundation

### 5.3.10. Scalability

Scalability is the applicability of a methodology to be used for various migration project sizes. Some methodologies can be suitable for moving large and complex workloads from traditional data centers to cloud infrastructures whilst others might be best suited for relatively partial and light migration. A migration methodology should be examined if it is appropriate to handle the intended scale of migration. This is done by firstly investigating activities that explicitly support scalability and characteristics that concerned with scalability (such as workload size, degree of interconnectivity of legacy applications and number of applications) and secondly, if the scalability of the methodology has been appraised in real-world scenarios. None of the reviewed approaches supports scalability nor refers to it. Since limited attention has been given to the scalability, future approaches should define the project size that are both appropriate and applicable.

### 5.3.11. Domain Applicability

A particular interest about the approaches is to understand the application domain for which they have been designed. Approaches were classified on the basis of their adherence to a particular domain. As a result, 18 domains were found, as shown in Table XV, whilst 23 approaches did not clarify the applicable area that they might suite.

Table XV. Application areas migration approaches

| Study | Domain |
|---|---|
| [S1], [S4], [S5], [S7], [S11], [S13], [S16], [S18], [S19], [S21], [S22], [S23], [S24], [S28], [S30], [S31], [S34], [S35], [S38], [S40], [S41], [S42], [S43] | Not specified |
| [S2] | Commercial Medical Application |
| [S3] | e-Tourism |
| [S6] | University |
| [S20] | Customer Relationship Management |
| [S8], [S39] | Open Source System |
| [S9] | e-Science applications (scientific workflow) |
| [S10] | Enterprise Resource Management |
| [S12], [S27] | Oil and gas |
| [S14] | Document Processing |
| [S15] | E-bank |
| [S17] | Document management systems |
| [S25] | Telecommunication |
| [S26] | ERP |
| [S29] | Medical Communications System |
| [S32] | Stock Market |
| [S33] | Digital Publishing |
| [S36] | e-Commerce |
| [S37] | Strategy Management Systems (SMS) based on Linux systems |

### 5.3.12. Tool Support

Developers can benefit from available or custom-made tools to carry out the migration activities. An approach can offer its tool or alternatively refer developers to existing third-party available tools in the cloud market. With respect to this criterion, 34 of 43 (79%) approaches did not offer any tools for carrying out migration activities. As tool support for the whole migration process may not be practical, approaches vary in their focus on automation support for migration activities and hence provide a partial support. As an example, the suite of tools (e.g. cost modeling, suitability analysis, energy consumption, and stakeholder impact analysis) offered by the model of Khajeh-Hosseini et al. [S6] are to aid an organisation to explore concerns in the early phase of the migration. Among them, the cost modeling tool is the most mature one which examines different deployment options of legacy application in the cloud. In general, 7 approaches provide tools for some activities in their suggested methodology.

Currently, two methodologies argue for inclusion of a wide range of tools for the entire migration process: (i) ARTIST methodology [S43] proposes Eclipse-based suites which are tightly integrated with its activities. Since produced work-products are stored in a shared repository, they can be accessed and modified other tools; and (ii) REMICS methodology [S26] that includes a set of tools that can be classified in the areas such as requirement management, knowledge recovery from legacy applications, re-transformation of application components to cloud architecture, and model-based testing. Finally, as far as availability of an approach's tool is concerned, it can be attributed as publicly available proprietary, or not specified. Table XII shows whether an approach fully or partially offer tools. Full description of the activities in this table is presented in Section 5.4.

Table XII. Approaches providing tool for migration activities

| Approach | Activity | Migration Type | Availability | Aim |
|---|---|---|---|---|
| [S6] | Context Analysis | V | Not-specified | Supporting decision making during the suitability analysis of migration in terms of operational cost, organisational change, energy consumption, and stakeholder analysis. |
| [S9] | Incompatibility Resolution | IV | Available | Supporting the data tier migrating to the public cloud and the refactoring of the application components. |
| [S11] | Design Cloud Solution | V | Not-specified | Allowing developers to model application components and define relevant criteria for the splitting components in the cloud. |
| [S39] | Legacy Application Understanding | All | Available | Extracting an ontology of source code, data, and architecture of the legacy application. |
| [S26] | Legacy Application Understanding | II | Available | Discovering legacy application components. |
|  | Resolving Incompatibilities |  |  | Resolving interoperability problem between legacy components and third-party cloud services. |
|  | Design Cloud Solution |  |  | Generating a platform specific implementation model from the cloud architecture model. |
|  | Test |  |  | Creating test cases and test scenarios. |
|  | Deployment |  |  | Generating a platform specific implementation model from the architecture model by taking into account the constraints relevant to the deployment of applications in the cloud. |
| [S40] | Elasticity (Resource Provisioning) |  |  | Configuring amazon services (e.g. storage, resource). |
|  | Resolving Incompatibilities | V | Available | Integration legacy components with amazon cloud services. |
|  | Network Configuration |  |  | Configuring network and dynamic resource provisioning (e.g. elasticity, virtualization, auto-scaling). |

| [S43] | Context Analysis | | | Analysing business and technical migration feasibility. |
|---|---|---|---|---|
| | Legacy Application Understanding | II | Available | Representing high-level and platform-specific models of the legacy application. |
| | Incompatibility Resolution | | | Transforming legacy codes to the cloud target platform. |
| | Test | | | Supporting functional and non-functional testing. |
| | Deployment | | | Deploying application components in the cloud. |

### 5.4 RQ2.2 What cloud-specific criteria are supported by these approaches?

The goal of the second dimension of the framework is to characterise existing approaches in terms of support for the various activities that might be required to perform in a migration lifecycle, and also to find existing gaps. Table G.1 (Appendix G) shows the extent to which each criterion is satisfied by each approach on the basis of three scale points including 'fully-supported', 'partially-supported', and 'not-supported' (See Table H1 in Appendix H). The rest of this section qualitatively elaborates scores in this table. We also highlight recommendations and advice offered by these approaches as a result of applying them in real scenarios.

#### 5.4.1 Migration Type

As mentioned in Section 2.1, there are several possibilities to make legacy applications cloud-enabled. This criterion indicates migration type for which an approach has been designed. Table XVIII shows the classification of approaches based on the migration variants. Along with the migration type, 18 out of 43 approaches define support for moving the whole application tier to the cloud, i.e. Type V. The idea is to encapsulate application into virtual machines, deploy in IaaS infrastructure, and add an additional layer to orchestrate the virtual machines. Twelve approaches provide support for moving data tier of applications to a cloud database solution in addition to the required adaptation activities to the application to access the data layer in the cloud (Type IV). As many as 13 and 7 approaches, respectively, propose activities for migration type I (i.e. deploying business logic of an application in the cloud) and type III (i.e. deploying a database in cloud infrastructure). Thirteen approaches focus on defining a series of activities to modify legacy components to make them cloud-enabled using existing application development APIs offered by PaaS or SaaS delivery models (Type II). Some studies did not define an approach for any particular migration type; rather they suggest an overall process model for migration. Twelve approaches were fallen in this group. Several migration types can be used to make an application cloud-enabled. For example, the business logic tier and data tier may be migrated to Google App Engine and Amazon Relational Database Cloud Service, respectively. This is why some approaches are repeated in the second row of Table XVIII.

Table XVIII Migration types and classification of approaches based on migration type

| Migration | Approach | Number |
|---|---|---|
| Type I | [S1], [S3], [S5], [S15], [S17], [S18], [S35] | 7 |
| Type II | [S1], [S5], [S8], [S12], [S13], [S15], [S17], [S18], [S26], [S34], [S35], [S41], [S43] | 13 |
| Type III | [S1], [S14], [S15], [S17], [S18], [S34], [S35] | 7 |
| Type IV | [S5], [S8], [S9], [S10], [S15], [S17], [S18], [S19], [S25], [S29], [S34], [S35] | 12 |
| Type V | [S6], [S8], [S11], [S15], [S16], [S18], [S22], [S23], [S25], [S28], [S29], [S30], [S33], [S34], [S35], [S36], [S40], [S42] | 18 |
| In general | [S2], [S4], [S7], [S20], [S21], [S24], [S27], [S31], [S32], [S37], [S38], [S39] | 12 |

#### 5.4.2 Unit of Migration

For many reasons such as regulations and privacy laws, it is unlikely that organisations migrate the whole of a legacy application stack to the cloud; and instead some components of the application are migrated whilst other are kept in local organisational network, and cloud services are offered to them. Hence, it deserves to investigate if an approach fit for moving a particular tier or whole application stack to the cloud. Given that, this criterion indicates which application tier is concerned for migration by the approach. Table XIX shows the result of classification of the approaches with respect to the unit of migration. According to this table, the majority of approaches focus on moving all the application tiers. Studies [S4], [S6], and [S38] view all the legacy assets in an organisation as a single unit for migration and propose a general approach for moving all the legacy applications to the cloud. Only study [S3] investigates running Web-Services, described using Web Service Business Process Execution Language, in the cloud via IaaS service delivery models.

Table XIX. Classification of approaches based on the unit of migration

| Unit of Migration | Study | Number |
|---|---|---|
| Organisation | [S4], [S6], [S38] | 3 |
| Whole Application Stack | [S1], [S2], [S5], [S7], [S8], [S11], [S12], [S13], [S14], [S15], [S16], [S17], [S18], [S21], [S20], [S22], [S23], [S24], [S25], [S27], [S28], [S29], [S30], [S31], [S32], [S33], [S34], [S35], [S36], [S37], [S39], [S40], [S41], [S42], [S43], [S44] | 36 |
| Data Tier | [S9], [S10], [S19] | 3 |
| Business Logic Tier | [S3] | 1 |

### 5.4.3 Analysing Context

Transition to the cloud is not merely a technological improvement of existing applications but also it is a substantial change in the way legacy applications hereinafter operate and are maintained. As an important first activity in a migration project, developers and organisations should conduct feasibility and business value analysis of the cloud as an IT strategic weapon, to empower legacy applications. In this regard, this criterion can be used to examine if there is a need to approaches that provide systematic activities for context analysis. In this survey, 27 approaches (62%) did not provide any suggestion on the context analysis, and 7 approaches (16%) only refer to this activity without any elaboration on how to perform it.

It was found that 9 approaches out of 43 (20%) explicitly support conducting context analysis. For example, Conway and Curry report their experience in validation of IVI Cloud Computing Life Cycle [S4], an Agile migration process model, within a number of organisations. They observed that *organizations that wanted to move to the cloud needed to fully understand the impact of the migration on the user community, and on IT support staff. Those organisations that did not understand this impact and that failed to plan accordingly either lost key resources or experienced resistance from the IT and user community – both during and after the migration*. In the Cloudstep, a model proposed by Beserra et. al. [S2], authors point *the idea [context analysis] is to anticipate the detection of potential organisational constraints that might affect the cloud migration decision, before carrying out any further analysis of the application itself*.

In the migration approach proposed by Khajeh-Hosseini and Greenwood [S6], the main focus is on the early phases of the migration process and it suggests three kinds of activities named as Technology Suitability Analysis, Energy Consumption Analysis, and Stakeholder Impact Analysis. They state that *understanding the significance and the extent of the organisational changes associated with cloud adoption is a difficult challenge. We argue that enterprises need to understand the breadth of changes and the effort required to make these changes in order to understand their benefits*. The ARTIST methodology [S43] defines an activity called business feasibility analysis of legacy migration which is to *provide not only economic information (ROI, payback, etc.) but also what are the main risks to be faced with the migration and the organizational processes affected by the uptake of the new business model*. Ahmad and Babar in their proposed framework [S21] stress two concerns during conducting context analysis, i.e. (i) determine the type of the application is to be migrated to the cloud since some applications may not benefit from the cloud such as safety-critical or embedded applications, (ii) effort and cost that required for the migration regarding perceived benefits. Table XX summarises the identified concerns from 43 reviewed approaches that should be taken into account when conducting context analysis. With respect to these concerns, the Cloudstep [S2] advises they are not mutually independent, rather their influence on each other should be investigated prior actual migration execution.

Table XX. Concerns should be investigated during context analysis activity

| Concern | [S2] | [S6] | [S4] | [S21] | [S27] | [S31] | [S23] | [S40] | [S43] |
|---|---|---|---|---|---|---|---|---|---|
| User resistance | √ | √ | √ | | | | √ | | |
| Loss of governance | √ | | | | | | | | |
| Dependency on legacy application | √ | | | √ | | | √ | | |
| Risk of unauthorised access | √ | | | | | | √ | √ | |
| Legal restriction | √ | | | | √ | | | | |
| Physical location of IT resources | √ | | | | | | | | |
| Energy consumption | | √ | | | | | | | |
| Variation on responsibilities | | √ | | | | | | | |
| Impact on organisational and daily activities | | √ | √ | | | | | | |
| Type of legacy application | | | | √ | √ | | √ | √ | |
| Required efforts and migration cost | | | | √ | | | | | |
| Scalability (workload fluctuation) | | | | | | √ | √ | | |
| Technical suitability | | √ | | | | | √ | | √ |
| Financial suitability | | | | | | √ | | √ | √ |

### 5.4.4 Understanding Legacy Application

It is common that the knowledge about legacy applications is often outdated, imperfect, and undocumented. Hence, it is required to identify any such characteristics of the legacy that may influence cloud migration. The activity of legacy application understanding aims at recapturing an abstract As-Is representation of application architecture in terms of functionality, different types of dependencies to other applications, interaction points and message follows between application components, and the quality of code blocks for reuse and adaptation. An architectural view of the application is produced as the output of this activity. Wu et. al. in [S33] report their experience on moving an online digital library search engine (CiteSeerX) to Amazon EC2 and found that a*n up-to-date and complete documentation can significantly reduce the length of learning and investigation time*.

The legacy understanding can be helpful in many ways: Some approaches such as [S12], [S21], and [S41] use produced high-level architecture models of the legacy application as a mean for automatic transformation to target cloud platforms. For example, CloudMIG [S41] proposes OMG's Knowledge Discovery Meta-Model (KDM) for extracting a meta-model of application and transforming it to a target cloud architecture. Based on the experience of moving legacy applications to IaaS, the approach of Sun and Li in [S16] incorporates legacy application understanding in order to get an estimation of required development effort for modernisation. In the approach by Rajaraajeswari et. al. [S27] the activity *Application Profiling* is to identify the application usage data which helps to understand the size of workload to migrate. This avoids from unexpected cost of cloud service usage when bills are issued. Specifically, it emphasises the following data to collect (i) CPU and memory usage (ii), storage data such as input/output operations per second, (iii) network data such as throughput and connections per second (iv) the node-level data to estimate number and type of machines are required when the application is migrated. As shown in Table XVII, 15 approaches (35%) explicitly define activities related to legacy application understanding, targeted for the cloud. However, 22 studies (51%) did not offer any activities or guidelines about the investigation of legacies. Table XXI summarises a collection of useful recommendations, each stresses a particular aspect of legacy application understanding, identified from approaches along with supportive technique to perform.

Table XXI. Summarised recommended activities for legacy application understanding

| Approach | Activity Name | Definition | Technique |
|---|---|---|---|
| [S1] | Recover activity area | Recover legacy application knowledge in a KDM format, which is as an intermediate representation supporting many programming languages, from which UML application models are generatable. | KDM |
| [S2] | Define application profile | Create a profile of legacy application including technical characteristic of the application that need to be considered for migration. | Not specified |
| [S7] | Discovery of application details from the source environment | Identify topology graph of enterprise's IT including legacy application and then all outgoing hosted-on, depends-on, and connects-to edges starting from the application's entry node. | Deep dive |
| [S12] | Architectural representation of the legacy application | Reconstruct an architecture model of the legacy application and any business entities and metadata. | Not specified |
| [S16] | Discovery | Collect configuration information of application including deployment model, a number of physical machines have been deployed in the local network, a network topology of these physical machines. Use this information to determine the deployment scale on the target cloud environment. | Not specified |
| [S21] | Architecture recovery and consistency conformance | Recover an architecture model of legacy source codes and check whether it is consistent with the source codes. This architecture model is necessary to find the design rationale that need to be preserved during modification/ transformation of application for the cloud. | Graph-based modeling |
| [S23] | Identify user load parameters | Discover and document groups of users, and the pattern of their use, teams of contractors in foreign countries, access time to the application, a particular time in the day or month in which application usage is heavy. | Not specified |
| [S24] | Describe existing capability | Analyse different available legacy artefacts such as source code, programming language, documents, users' knowledge, configuration files, execution logs and traces, integration with third party offerings, dependencies, database, upgrades procedures to identify artefacts that are related to be considered migration project. | Top-down approach and bottom-up approach |
| [S26] | Recover | Model "as-is" and "to-be" architecture and deployment models. | KDM |
| [S27] | Application dependency mapping profiling | Identify dependencies among applications in the local infrastructure and their usage data. This helps estimating the size of migration to the cloud. Based on the node on which the application is executed, the following data need to be identified: CPU and memory usage, data storage, network latency, and input/output operations per second, network data. | Not specified |
| [S37] | Discover | Identify static aspects such as source configuration details, network IP, and operating system type and dynamic aspect such as CPU usage and dynamic workload characteristics of application. | Not specified |
| [S39] | Capture | Identify key concepts and relations in the legacy application. | Ontology |
| [S41] | Extraction | Extract a utilisation model of legacy application including statistical properties concerning user behavior like service invocation rates over time or average submitted datagram sizes per request. | KDM |
| [S42] | Initial screening and analysis | Collect key data on existing workloads, applications and their dependencies, server utilisation, machine and operating system type, hardware and software characteristics. | Not specified |
| [S43] | Application discovery & understanding | Extract an overall logic of the legacy application and any different available legacy artefacts relevant for the migration scenario. | Model transformation |

### 5.4.5 Analysing Cloud Migration Requirements and Objectives

As the name implies, approaches may define some activities to specify and model the functionalities required to be fulfilled through legacy application enablement. Examining this criterion gives an insight whether developers seek for approaches that systematically define activities to capture and analyse requirements and ensure migrated application addresses the requirements of users. Common requirements engineering techniques (.e.g. interview, prototyping, and workshop) still are useable to elicit and analysis requirements from users, developers, and managers as it can be seen in approaches [S8] and [S26]. However, the requirement analysis in the context of cloud migration is with a specific focus on elasticity and scalability application requirements [S18], computing requirements [S19], inter-operability requirements for deployment in the cloud [S21], security and regulatory requirements [S23], and storage space requirements in the cloud [S33]. According to Table XVII, in 15 (35%) of reviewed approaches, it was found that they define related activities for the identification of expected requirements. However, 7 (16%) and 26 (60%), respectively, partially or does not define any activity for requirement identification and analysis.

### 5.4.6 Planning Migration

Once cloud migration is found as a feasible decision, a plan is defined that guides the rest of migration process. Feedback from stakeholders is analysed to define a migration plan. Support for activities related to the planning has been included in 12 reviewed approaches (28%).

According to [S27], information obtained from the activity legacy of application understanding is an input source for the planning. For instance, if legacies X and Y are using the same database server, a possible migration plan would be a combined move or splitting the dependencies.

On the other hand, some approaches extend planning with a specific focus on cloud characteristics. With respect to this, Strauch, S., et al. propose a methodology for moving e-science applications to the cloud by acquiring services from Amazon AWS and Microsoft Windows Azure [S9]. Planning in this methodology is defined as actions in order to resolve potential incompatibilities between legacy components and target cloud database solution without modifying the application business logic. Likewise, Pahl et. al. consider planning in terms of cloud service provider capabilities, addressing contract with partners, distribution of project team, capabilities of migration team (e.g. technology, skills, and tools), and defining metrics and milestones [S18]. In the process model of Legacy-to-Cloud Migration Horseshoe, proposed by [S21], authors incorporate the influence of cloud provider selection and migration type as main factors to develop a migration plan.

Furthermore, defining a proper roll back plan, i.e. switching to the previous version of the application at any stage of migration reduces the risk and exposure to organisation business. With respect to this, the approach [S4] highlights that *the option to roll back to an in-house version at any stage significantly reduced the risk and exposure to the business. Organisations that experienced difficulties in the transition to cloud computing missed vital steps in their planning*. Similarly, [S33] recommends defining *Backward Availability* for critical migration projects in the case new cloud application fails.

### 5.4.7 Cloud Service/Platform Selection

Approaches can be assessed based on the extent to which they properly define activities to identify, evaluate, and select a set of cloud providers that might suit organization and application requirements. In justifying this criterion, Chauhan and Babar examined their suggested process model for moving an open source system to Amazon Web Service and Google App Engine [S8]. They found that a better alignment between the quality attributes of application and cloud service provider can make the migration process quite easy. They conclude that suitability of cloud environments in term of security and sensitivity is pivotal in the success of application migration the cloud. Similarly, from a cost perspective, Tran et. al. [S5] report a breakdown of activities involved in moving a .NET n-tier application to run on Windows Azure and highlight efforts required for deciding cloud providers is major. They underline the influences of selecting cloud platforms on the required effort for the rest of migration process. If the selected cloud platform technology is highly similar to the legacy, less modification in the legacy is required.

In general, from 43 reviewed approaches the following important factors were identified for investigation when conducting the activity cloud provider selection: *variety of service models offered by provider*, *price model*, *the form of SLA*, *additional services such as backup, monitoring, auto-scaling*, *security mechanisms*, *implementation technologies which are supported by provider such as programming languages, development platforms*, *allowing access to internal operational logs*, *physical location of application data* [S2], *sustainability* [S4], *commitment durability with cloud provider*, *business objectives* [S8], *degree of automation, storage encryption mechanisms*, *storage format and exchange*, *developer SDKs* [S9] and [S10], *required configuration effort*, *deployment speed*, *consistent development experience* [S17], and *traffic bandwidth at the client network* [S29].

Once a provider(s) is selected, a contract between the legacy application owner and the cloud service provider is signed. The application owner needs to specify the scope of expected service provisioning by cloud provider prior migrating application components to the infrastructure of a cloud provider. The negotiation is finalised with producing an SLA which specifies the boundary of usage and provision of cloud services is produced. According to IVI lifecycle model [S4], *building a relationship with the cloud supplier was the key to success in many of the projects we studied*. As mentioned in Section 2.3, cloud environments can raise application licensing issue meaning that by running the application in the cloud, many instances can be dynamically available. From the reviewed studies, 4 approaches incorporate addressing licensing issues. For example, Amazon methodology [S40] recommends that for the application that organisation is tightly coupled with complex third parties application, which have not been migrated to the cloud, developers should extend the legacy application with new components (e.g. VPN tunnel) so that cloud services can be indirectly offered to them. [S23] suggests enabling license tracking through monitoring connections between the application and cloud resources.

### 5.4.8 Re-Architecting Legacy Application

The aim of this activity is to evolve the legacy architecture towards a new architecture which addresses cloud-specific characteristics. It uses legacy architecture model and requirement model which are outputs from activities legacy application understanding and analysing requirements, respectively. When it comes to the re-architecting legacy application for the cloud, several sub-activities might be required to carry out as described below.

**(i) Define Cloud Architecture Model.** One important aspect of re-architecting is to find suitable components for migration and re-arrangement of their deployment topology in the cloud environments. With respect to this, approaches vary in their support for this sub-activity. For example, Conway et. al. [S4] states *this [component selection] will require an understanding of the current state, so that it can be compared to the desired future state*. Andrikopoulos et al. [S35] numerate several factors are taken into account for component selection such as data privacy, expected workload profile, acceptable network latency and performance variability, availability zone of cloud providers, the affinity of components in the cloud, and the geographical location of cloud servers. Among the reviewed approaches, Leymann et al. [S11] proposes a concrete technique for component selection and distribution in the cloud via transforming application architecture into a graph partitioning problem and using existing algorithms such as simulated annealing problem to optimise the distribution of components among different cloud nodes.

**(ii) Enabling Application Elasticity.** Running applications in the cloud does not help to resource efficiency and scalability issues by itself. Rather applications should be able to grow and shrink resources under dynamic workload and maximise resource efficiency (Vaquero et al., 2011). Elasticity is triggered via a set of rule/conditions related to specific workload or threshold, events, or metrics. Often legacy applications have not been designed with a support for dynamic resource provisioning in mind and hence adaptation is required to enable this feature. Elasticity can be implemented by developing a new component either embedded in the application or separately hosted on a server node in order to continuously monitor the application/component resource usage variables and then performs appropriate action based on scaling rules.

Only Ridha et. al.'s approach introduced in [S22] defines generic activities for deploying existing applications which assure elasticity. It consists of three activities: (i) describing required computing resources, the life cycle of the application and its offered services, and scaling rules of the application and its services, (ii) automatically provisioning resources from selected clouds for the application, and (iii) monitoring and ensuring application performance. Two approaches rather provide high-level advice: Andrikopoulos et. al. [S35] point out aspects that needed to be considered during enabling elasticity in an application in terms of *what to scale*, *how to scale*, *when to scale*, *which scaling strategy to be used*, and *scaling latency*. Likewise, with respect to the approach of Maenhaut et al. [S29], to handle elasticity it is required that application components be decoupled and communicate in a-synchronised manner.

**(iii) Enabling Multi-Tenancy**. According to (Bezemer and Zaidman, 2010, Guo et al., 2007), re-architecting of a legacy application to support multi-tenancy includes the following aspects:

— The first aspect is *security isolation*. Enabling multi-tenant applications, specifically in migration type II, raises security risk as different tenants use the same database and running application instances. It is necessary to assure each tenant can only access to its data and to be protected from unauthorised access by other tenants which are running in the same cloud. As off-the-shelf database management systems might not support multi-tenancy (Jacobs and Aulbach, 2007), securing the data tier of application should be properly addressed. Furthermore, the code blocks of the application reflecting organizational business processes (.e.g. BPEL processes) might need to be secured prior deploying in the cloud to protect from unauthorized access by other tenants. Hence, assuring confidentiality of code execution, for example through encryption mechanisms, in the sense that no other tenants will be able to access, read, or alter the code blocks within the running application instance in the cloud.
— The second aspect is *customisability*. Each tenant may have a wide range of functional and non-functional requirements. For example, one tenant might want to have the ability to customise application user interface whilst other one need to change the sequence of application workflow and code customisation. Customisations are applied by a tenant should not affect other tenants (e.g. application availability). To achieve this, configuration points, in the form of application template, should be defined in the code blocks of the application.
— The third aspect is *fault isolation* meaning that if a fault occurs in the same instance of a running application instance being used by multiple tenants in a shared manner; it should not be propagated to other tenants using that instance. The application should monitor its internal state to detect the faults, prevent its propagation, and repair it in a timely manner.
— The fourth aspect is *performance isolation* which is to guarantee the performance of one tenant from the negatively being affected by the performance usage of other tenants in unforeseen behaviors. If this is satisfied in all situations, then the application is performance-isolated.

Given the above aspects of multi-tenancy, only the approach proposed by Maenhaut et. al. [S29] incorporates explicit activities to add customisability and security isolation to legacy applications are to move to a hybrid or public cloud environments. That is, enabling multi-tenancy requires the following steps: (i) decoupling databases, (ii) adding tenant configuration databases, (iii) providing tenant configuration interface, (iv) dynamic feature selection, (v) managing tenant data, users, and roles, and (vi) mitigating security risks.

Andrikopoulos et. al. [S35] point the concept *multi-tenant aware applications* and recommend addressing two aspects of multi-tenancy as (i) supporting message exchange isolation per tenant, and (ii) administrating and configuring the application per tenant. Other approaches only recommend considering multi-tenancy activity without any proposal on required activity to perform.

**(iv) Incompatibility Resolution.** If legacy application components are to be bound to specific cloud services, potential incompatibilities between the legacy application and these services should be identified and accordingly resolved through adaptation mechanisms. With respect to the existing approaches, three forms of adaptations might be required to resolve incompatibilities.

— *Code refactoring.* A basic form of incompatibility resolution is through modifying legacy codes so that they can interact with the cloud services. This can be the modification of legacy application interfaces in order to remove mismatches (e.g. interface signature, operation ordering, operation names, message format, interaction protocols, and data type) with cloud service interfaces. Also, behavioral adaptation is performed if there is a mismatch in the granularity of interaction messages between application components and cloud services. An explicit support for code adaptation was found in 11 approaches, i.e. [S3], [S5], [S8], [S9], [S10], [S12], [S14], [S15], [S25], [S35], and [S43]. Other approaches such as [S7], [S13], [S17], [S20], [S21], [S24], [S26], [S29], [S30], [S31], [S33], [S39], [S40], [S41], and [S42] only confine to general advice on code adaptation.

— *Developing Integrators/Adaptors.* If the code refactoring, as mentioned in the previous item, is costly, then developing integrators/adaptors can be served as an alternative solution to hide incompatibilities. Adaptors provide an abstraction layer, keeping the application code untouched and facilitating application interoperability and portability. With this regard, the ADM (Architecture-Driven Modernisation) approach of Zhang [S12] suggests developing wrapper layers (e.g. Web services) on legacy components. Some approaches include high-level classifications of the possible adaptors that developers should build. In this way, CMotion approach [S30] defines two types of integrators focusing on aspects: *when* and *how*. The former is to define adaptors for transformation application components before deployment in the cloud whilst the latter is to define ways of applying changes, i.e. automatic vs. manual adaptors. A similar view has been also defined by Miranda et. al. [S14]: static (or design-time) adaptation, which includes the adaptations required to be performed prior the application is running; and dynamic (or run-time) adaptations at the runtime application execution in the cloud. Jamshidi in [20] presents a catalogue of patterns to integrate legacy components with different cloud platforms.

— *Data Adaptation.* Migrating the data tier to a cloud database solution can also cause different mismatches. For instance, although Cloud database solution SQL Azure is similar to traditional SQL Server, distributed transactions are not supported by SQL Azure as SQL Server does. This may imply data type conversions, query transformation, database schema transformation, and developing runtime emulators. The methodology suggested by Strauch et. al. [S9] and [S10] provide a richest of *Cloud Data Patterns* in order to address incompatibility issues between traditional databases and cloud database solutions. This patterns are classified in three groups as *functional patterns* (e.g. data store functionality extension, emulator of stored, procedures, local database proxy), non-functional patterns (e.g. local sharding-based router), and confidentiality patterns (e.g. confidentiality level data aggregator, confidentiality level, data splitter, filter of critical data, pseudonymizer of critical data, anonymizer of critical data). Additionally, the methodology suggested by Oracle [S19] defines the following activities for database migration (i) database schema migration (involves migrating tables, indexes, and views in a database), (ii) data migration, (iii) database stored program migration triggers, and views, (iv) application migration, and (v) database administration script migration.

Table XXII summarises the list of identified mechanisms proposed by the existing approaches for developing integrators and data adaptors.

Table XXII. Summarised of resolution mechanisms to resolve incompatibilities

| Study | Mechanism | Legacy Tier | Definition |
|---|---|---|---|
| [S9] | Data store functionality extension | Data Tier | As cloud database solution may not support functionalities that legacy data tier offers (for example stored procedures), the data tier is extended with |
| [S9] | Local database proxy | Data Tier | Cloud data store might not offer horizontal scalability for data reads. To address this feature, the cloud data store is configured through applying a single master/multiple slave model where the master manages writing data and slaves are replicas for read operations. All data traversing from each data tier is conducted through a proxy. |
| [S12] | Web-Service generation | Application Stack | |
| [S14] | Adapter generation and injection | Business Tier | The adaptor consists of a proxy with the concrete implementation of the service interface and is injected into the application as an instance of the component required for consuming the selected cloud service. |
| [S15] | Service adapters | Application Stack | Developers should identify adaptation points in the application and define intermediate artefacts which automatically injected between interaction between legacy components and cloud services and address incompatibilities. |
| [S30] | Adaptor | Application Stack | Adaptors are classified into two dimensions: when and how. The when dimension denotes times that adapters are applied. The how dimension represents how adaptors apply their changes. |
| [S20] | Hybrid refactor with cloud adaptation | Application Stack | A component adapter (e.g., legacy façade) is adopted to provide integration of the legacy components with re-hosted cloud-based components. |

**(vi) Applying Architecture Design Principles.** From the reviewed approaches, a range of recommended application design principles was identified which need to be considered by developers when re-architecting a legacy application to target cloud. According to [S8], [S14], [S17], [S22], [S34], [S36], and [S42] these principles are as follows:

—*Application decoupling.* To take the advantage of dynamic deployment, one architectural requirement is to decouple application components so that they can interact in a transparent manner. Decoupling also enables independent elastic scaling of the components (i.e. dynamically adding/removing more instances of the same components), minimise time required for code refactoring and test in the case of changing cloud provider, handle a-synchronised communication, as well as simplify coping with the components failures.

— *Stateless Programming*. To support dynamic deployment and independent component scalability, components should be stateless and minimise storing contextual data. Moreover, the coexistence of multiple running instances of the same components in the cloud environment requires a new session management mechanisms in order to track tenant's behaviors and ensure their security when a number of instances is increased or decreased based on the server workload.

—*Transient Fault Handling.* Performance variability and network latency can have an impact on QoS of migrated application. Developers should implement mechanisms to detect and handle transient faults that occur in cloud environments.

Again it should be noted that performing the above-mentioned sub-activities mainly depends on selected migration type. Given definitions of migration type variants in Section 2.1, Table XXIII shows the situations in which a sub-activity is required to perform.

Table XXIII. Sub-activities for re-architecting and migration types that trigger them

| Re-architecting sub-activity | | Type I | Type II | Type III | Type IV | Type V |
|---|---|---|---|---|---|---|
| Component selection and distribution | | Very Likely | Very Likely | Very likely | Very likely | Very likely |
| Enabling Elasticity | | Not Likely | Very Likely | Not Likely | Not Likely | Very likely |
| Enabling Multi-Tenancy | | Very Likely | Very Likely | Not Likely | Not Likely | Very likely |
| Incompatibility Resolution | Code adaptation | Very Likely | Very Likely | Very likely | Very likely | Maybe |
| | Developing Integrators/Adaptors | Somewhat Likely | Somewhat Likely | Not Likely | Very likely | Not Likely |
| | Data Adaptation | Not Likely | Somewhat Likely | Somewhat Likely | Very Likely | Somewhat Likely |
| Applying Architecture Design Principles | Application decoupling | Very Likely | Very Likely | Very Likely | Somewhat Likely | Somewhat Likely |
|  | Stateless programming | Very Likely | Very Likely | Somewhat Likely | Somewhat Likely | Very Likely |
|  | Transient fault handling | Very Likely | Very Likely | Very Likely | Very Likely | Very Likely |

### 5.4.9 Training

As cloud platforms come with a particular set of tools and APIs, training that covers new cloud services is necessary. Some existing approaches recommend incorporating training activities for IT staff (developers and managers) as a part of a migration process. At the managerial level, for example, there is a change from a traditional licensing model to pay-as-you-go or post-usage billing [S18]. On the other hand, developers need training on new programming concepts such as asynchronous interaction, distributed state and session management, caching, scale out across data centers and providers (scalability), multi-tenancy [S35]. Tran et. al. state that training activities should be incorporated into migration process since all cloud services may not support some features offered by legacy technologies [S5]. Sun et. al. [S16] put one step further and define a process-based effort estimation approach metrics to measure the capability of a development team on the basis of mastery of developers on conducting migration activities.

### 5.4.10 Test and Continuous Integration

Once adaptations applied to the application test activity is performed to ensure that application conforms to the expectation of the cloud migration. It is an unavoidable activity once adaptations applied to the application. Like tradition software development, the test activity includes testing both functional and non-functional aspects. However, in the context of cloud migration, various cloud-specific tests are to perform including security test, interoperability test, and workload test. Agile REMICS [S1] defines a phase called validation activity which, ,consists of the following steps for migration type V: (i) define testing infrastructure, (ii) identify and refine requirements to be tested, (iii) generate acceptance testing, (iv) import models elements to be tested, (v) define testing procedures, (vi) implement testing strategy.

Other approaches, however, provide general advice on conducting tests as described in the followings. For many reasons such as workload or user preference, a running application may move from a cloud environment to another one. With respect to this, a lesson learned from applying the migration framework [S15] is to test application for interoperability in heterogeneous cloud environments. Therefore, a certain level of integration and interoperability testing is required. Important tests that should be taken into account are *Data Verification*, *Test Backup and Recovery Plan* before they required [S23], *Test Network Connectivity* (connection between cloud services and local network), *Test Connection Speed* as there is network latency to receive a response from the cloud server located in different geographical areas [S33] [S40], *test provider performance variability* and *test application latency* due to the network performance variability [S35].

An important aspect of the test is to address continuous integration. Continuous integration is to add a new increment to all tier of the application as well as any host in a smooth and coordinated manner. Cloud provides a new environment for automated testing by adding and removing servers for test and integrating new increments to the working application. This reduces costs of deploying, maintaining, and licensing internal testing tools. Except for [S33], none of the studies provide support for continuous integration. Wu et. al. in [S33] recommend a technical mechanism called *staging and production* in order to use separate servers for test activity. In the staging environment, developers can test the application within a production-like environment (.e.g. infrastructure configuration). That is, a deployment environment in which all the tests to be performed in order to detect bugs, performance and other issues before the application can be used by users. All releases of application are tested first on the staging environment. Once, the application completely implemented and tested, it is pushed to the target environment, i.e. production, as an increment of the application.

### 5.4.11 Environment Configuration

Once the application migrated to the cloud, the connection between the local network and the application in the cloud is still required. Ten out of 43 approaches ([S3], [S5], [S9], [S10], [S12], [S19], [S23], [S28], [S40], and [S42]) have provided support for adjusting application environment configuration for the target environment. With respect to these approaches, the environment configuration embraces the following sub-activities:

—Reconfigure organisation network setting such as ports, firewalls, and anti-virus, reachability policies, and connection strings to the database.
—Give privileges to application tenants/users to assure the security of application still is satisfied in the cloud environment.
—Create installation scripts and setup different third-party libraries and tools which may be used for monitoring and reporting runtime application behavior, though this needs less effort if cloud provider has already done it.

### 5.4.12 Continuous Monitoring

The dynamic and unpredictable nature of the cloud environments necessitate continuous monitoring of application and cloud resources to assure successful SLAs (e.g. [S1], [S4], [S22], [S23], [S26], [S28], [S33], [S38], [S40], and [S43]) as described in followings:

- *Measuring*. Collect critical data about the application health and performance, traffic patterns, vital signs such as CPU and memory usage, network traffic, and disk usage. This data are used to detect deviations from SLA and runtime application adaptation and optimisation. Measuring can be fulfilled either by developing a new component which is integrated with the application and is responsible for this purpose or can be offered by cloud service provider [S1] [S4] [S22] [S23], [S28] [S33] [S40] [S43].
- *Updating patches and periodical backup.* During the application operation in the cloud, regular patch updated and database backup needs to be addressed either by application owner or by the cloud provider [S28] [S33] [S40].
- *Metering and managing bills.* There are two aspects in this. Firstly, billing management is required to support by the application owner where application offers services to tenants. A new component might be added to the application which tracks the tenants using the services and bill them accordingly. Secondly, the application owner is billed for the services (e.g. infrastructure) used for developing and maintaining the application. Hence, the owner should monitor resource usage, track logs periodically, and identify lifecycle patterns of each application instance, and detect suspicious high resource usage which can inflate the cloud service usage fees [S38] [S43].
- *Replicating and synchronising.* Replication is performed to support high business continuity and minimal downtime by running several application instances on a variable number of cloud infrastructures. The partitioning and replication may entail extending the application for support of synchronisation between the components (local replicas and those hosted in the cloud), though if a cloud provider supports this, then the application can leverage it [S33].
- *Terminating idle instances.* Removing under-utilised application instances to increase utilisation of the overall application [S40].
- *Decommission legacy components.* Once the application becomes operational in the cloud, the old legacy components may no longer be required and hence they are retired and the assigned resources to them are released [S23].
- *Withdrawing.* In some cases, the application owner wants to withdraw the application from the cloud. If required, the application is removed and other acquired resources (e.g. data storage, CPU, and network bandwidth) are released to reduce costs [S26], [S28].

# 6 CHALLENGES AND FUTURE DIRECTIONS

It is not feasible for an approach to cover all the criteria; rather, it may focus on a subset of them based on its objective. Therefore, it was hard to rank approaches from lower priority to higher. The evaluation framework can be viewed as an indicator of the primary focus of an approach instead of judgment on its completeness. In other words, if an approach does not support a certain criterion, it does not necessarily mean that it absolutely cannot be useful. Instead, it means that criterion is not its main focus.

While all the reviewed approaches in the literature have merit and form a rich source of necessary activities and recommendations to be learned, our deep analysis revealed that still there are challenges which are yet to address. Given the RQ2 and in the light of evaluation results in Section 5, overcoming the following challenges may open new possibilities to ameliorate the state of process models for cloud migration in the literature that constitute future research directions in the field.

— **Approaches suffer from a sound research quality.** Like other fields of software engineering, a rigorous research methodical is of utmost significance in order to advance and better understanding the field. With respect to the first dimension of the evaluation framework, only 6 out of 43 (14%) included a section in the paper to explain their research methodology. This lack is also true for the data collection (7 approaches met this criteria) and analysis techniques (4 approaches only) when applying and validating approaches. This undermines the trustworthiness of the existing approaches. We believe that using proper research methodologies will improve the quality of individual approaches and enable to combine their adoption results. A good example of applying of a sound research methodology in the field of legacy migration can be seen in the study by Razavin (Razavian, 2013) where she applied an exploratory action research in order to get a deep understanding of what academia and practitioners perceive about the migration process of legacy to SOA and identified the categories of activities that are carried out during migration. We believe a similar exercise can also be equally applied in the field of cloud migration.

— **Approaches are not tailorable.** It is highly recommended that methodologies should be tailored before use (Sommerville and Ransom, 2005). This strand continues where cloud migration is concerned. For example, one study suggests: *There is an immense need to identify correct process model for the deployment of cloud-centric environments in order to meet changed business requirement of clients. One solution can never fit all problems; likewise, there is a need of customised cloud for individual businesses and dynamically changed requirements of clients* (Z.Mahmood, 2013). With respect to the criterion tailorability, as defined in the first dimension of the evaluation framework, any existing approach adds a configurable aspect to tailor it for a migration project at hand. From the reviewed approaches, only 5 approaches argue a need for the approach tailoring but without proposing any mean to do that (Appendix E). To address approach tailoring, one can apply a method engineering approach (Brinkkemper, 1996), that is based on this thought that instead of looking for a universal software development methodology, developers should construct a new bespoke methodology by using existing *method fragments* that are stored in a *method library* or *repository* so that the produced methodology meet project characteristics. Applied to the field of cloud migration, it means as developing a repository of migration fragments along with mechanisms to construct situation-specific methodologies through assembling these fragments which fit project at hand. In the context of SOA, Börner (Börner, 2010) and Khadka (Khadka et al., 2011), respectively, proposed method engineering approaches for construction situational service identification and methodology for legacy to SOA migration. We believe this kind of research can be applied in the field of cloud migration.

— **Lack of adequate support for multi-tenancy, elasticity, test and continuous integration**. One interesting observation from reviewing 43 approaches is that only [S35] covers multi-tenancy merely from the customisation aspect. Elasticity has been also supported only by one approach [S22]. Continuous integration is partially supported by two approaches [S23] and [S32]. It seems that there is a clear lack of support in the literature when it comes to evaluate existing approaches with respect to these activities as they have defined very broad and inarticulate hence got the got the lowest overall score in this review. Over the cloud computing literature, one can find many ad-hoc techniques on these activities. Future research required to properly abstract and structure these techniques in the form of activities (or method fragments) and incorporate them into existing migration approaches migration lifecycle.

— **Developer roles have not been sufficiently honed.** The definition and responsibilities of roles involved in a migration process have not been well described in existing approaches. Specifying roles will make clear for developers their exact responsibilities and activities to the roles and may lead to better governing migration process. Besides a need for the elaboration of different roles and responsibilities, exiting approaches do not incorporate cloud-specific challenges into the role definition. Rather they use the common SOA roles (.i.e. service provider, consumer, and broker). However, an overview on blogs and white papers in online cloud communities shows that new development roles and skills should be considered. Stafford, who is executive editor of TechTarget, gives the some examples of new roles such as *continuous integration skills for real-time testing and diagnostics*, *virtual infrastructure configuration*, *Hadoop on cloud for handling big legacy data* (Stafford, 2013). More research on necessary roles should be conducted to characterise not simply developer roles but also any stakeholders involved in cloud migration. Gu and Lago proposed a role-driven migration process model for service-based application development (Gu and Lago, 2007). The model has two dimensions where a horisontal view explicitly shows the activities associated with roles in SOA and vertical view shows the interaction and cooperation between the roles. This can be a good starting point to do a similar research about roles in the context of cloud migration lifecycle and unfold cloud-specific roles.

— **Lack of a unified process model for cloud migration.** Review of the existing approaches shows that they use different terms and definitions to describe same constructs in a migration process. With many advantages that variety of cloud migration approaches offers, nevertheless, it would be beneficial to have an overarching view of legacy to cloud migration process. Furthermore, we observed that the approaches are often combined with technical-centric concepts which are often not homogenous and sometimes limited to certain cloud-specific platforms. Irrespective of technical aspects of cloud migration, the question here is how would developers grasp a quick and platform-agnostic view of existing cloud migration processes? Synthesising a unified and well-abstract cloud migration process model from existing approaches, would be advantageous in terms of facilitating understanding of cloud migration process, lucid knowledge transfer across the community of cloud researchers and practitioners, interoperability of cloud migration methodologies across process modelling tools, as well as specialising different parts of it independently. The need for unified migration model models has been corroborated by cloud researchers (Zimmermann et al., 2012, Hamdaqa and Tahvildari, 2012) but still no work with this focus exists in the literature. The reference process models for cloud migration are still a gap in the literature and worthwhile pursuit.

— **Other problems.** Beyond the above-mentioned problems, the criteria traceability, scalability, and formality have been weakly supported by existing approaches. Furthermore, automation support in performing migration activities has not been sufficiently supported.

**7 THREAT TO VALIDIDTY**

There are two limitations in the survey presented in this paper: bias in publication selection and imprecise data collection.

Firstly, we focused on studies whose main objective was to suggest an approach (e.g. methodology, process model) for the legacy to cloud migration, specifically in the form of a process model. This resulted in identification of 13 primary approaches in the literature. Through snowballing technique, (searching for all references contained in the studies), we found that studies do not necessarily use the terms process model and methodology consistently; and even in some cases, they combined these with other objectives that results in ambiguity in describing aspects of migration process. For example, [S35] was a well-cited source and included very useful practice and considerations for the legacy to cloud migration; however, this case was not identified through our initial search strings. To reduce the likelihood of missing relevant papers, firstly we broadened the search strings, as shown in Table III in Appendix A, to cover more relevant papers; and secondly, we performed an extensive manual search in scientific databases and conference proceedings as enumerated in Appendix A. However, it is still possible that some related work may have been missed.

Secondly, although we have used the evaluation framework as a lens to extract data items from the approach, still we believe our data extraction can be imprecise since in some cases we found that approaches did not clearly describe their suggested migration model. We frequently found that research methodology of approaches including validation techniques, contextual information, and data analysis had not been properly explained. This could affect the quality of our data extraction. Thirdly, the evaluation results reported in this survey have been based on the available documents on the approaches. A more realistic evaluation could be performed through applying approaches in real-world migration scenarios. However, such evaluation is out of the scope of this paper.

Finally, we acknowledge that despite conducting a rigorous review procedure for the ratings, still the rates rely on subjective interpretation of the authors, hence the risk of bias is probable due to the interpretivist and subjective nature of any survey paper. To alleviate this threat, it should be noted that the rating was not performed in a single step; rather, it was made after a few refinements, agreed collectively by all four authors. More specifically, each identified paper was read and immersed within the mind of researchers, as recommended by (Braun and Clarke, 2006) and then compared with other identified papers. This helped us to get familiar with the depth of each study. In addition, in order to rate the approaches in a precise manner, we emphasised on a proper extraction of all relevant constructs (e.g. phases, activities, work-products) from the identified 43 studies as a basis for the rating and with respect to the defined criteria. For the extraction of the constructs, the authors were actively involved for identification of the constructs.

## 8 CONCLUSION

We presented a systematic literature review on the legacy to the cloud migration from the process model perspective. As far as RQ1 is concerned, we reviewed, evaluated, and characterised existing proposed approaches suggesting a methodological solution for moving legacy applications to cloud environments. This was carried out via conducting an exhaustive systematic literature review including manual and automatic search on major electronic scientific databases since 2007, screening many papers and eventually leading to 43 papers related to the migration legacy applications to cloud environments. An evaluation framework was suggested to highlight important features, activities, and recommendations of existing approaches. A particular strength of the framework is its potential application as a yardstick in selecting approaches or a subset of them as to satisfy specific requirements of a given cloud migration scenario at hand.

Regarding RQ2, several few research opportunities discussed in this paper. Firstly, we found that researchers have not employed significant research approaches such as interviews, focus groups, observations, surveys, design efforts, and archival materials so as design and evaluation migration approaches. While the cumulative learning and understanding of cloud migration can be attained via sound research methodologies, this essence has been less incorporated in the developing of current approaches by approaches' authors. Secondly, a central aspect of the cloud migration is that a fixed migration approach is not suitable for all migration situations. Nevertheless, the literature review revealed that little work exists that provides a mean to design situation-specific approaches with respect to migration project context and goals have been largely ignored in the literature. In addressing this gap, we suggested engaging the engineering approaches to identify method fragments from existing approaches and literature and combine them to design migration approaches fit a migration scenario at hand. Thirdly, existing approaches need to be empowered to with relevant method fragments to the aspects multi-tenancy, elasticity, test, and continuous integration which have been weekly supported. Fourthly, the definition of new roles specific to cloud application development has not been explored in existing approaches. As mentioned in Section 5.3.4, a methodology should specify producers that are involved in the cloud migration process. Fifthly, recognising that there is a sheer volume of cloud migration research, which is currently dispersed and fragmented, implies a need for a generic reference model aiming at integrating existing literature. The fact that each year a considerable number of research papers are published in the field cloud computing, where each reports different solutions, experience reports, and recommendations to move legacy assets to cloud environments, itself is an evidence that the field has reached a maturity point where the development of such a generic reference model is mandatory. Sixthly, some criteria such as traceability, scalability, formality, and automation have been not been properly supported by existing approaches. The current state definitely calls for further enhancement of the cloud migration with more methodological approaches.

Finally, the results of analysing existing approaches have made a rich inventory of important activities, recommendations, and concerns in the existing cloud migration approaches that are commonly involved in the migration process in one place as described Section 5.3 and 5.4. In our view, this inventory is a great contribution of this work which can be used by academia and practitioners to understand the essence of the cloud migration process. In addition, the results of the evaluation existing approaches and the proposed evaluation framework is helpful for practitioners to get an understanding of the applicability of each individual approach and on the other hand comparing of the approaches to select one or a subset of them with respect to their migration project requirements. In this regard, the survey presents a basis for a well-informed decision. Another key contribution of this survey is to enable people in the cloud computing community to get an overall view of the current state of research to the methodological aspect of legacy application migration to cloud environment including key concerns, activities, and criteria need to be taken into account. It is enjoyed by a novice who will engage in the cloud migration research and anyone who is interested in the methodological aspect of moving legacy enterprise applications to cloud environments. We hope that they expand cloud computing body of knowledge by addressing the identified challenges.

**ACKNOWLEDGMENTS**

The authors would like to thank Professor Fethi Rabhi for his constructive comments and suggestions on improving this paper. They also wish to acknowledge comments made by two anonymous reviewers that greatly strengthened the manuscript.

**REFERENCES**

ANDRIKOPOULOS, V., BINZ, T., LEYMANN, F. & STRAUCH, S. 2013. How to adapt applications for the Cloud environment. *Computing,* 95**,** 493-535.

ARMBRUST, M., FOX, A., GRIFFITH, R., JOSEPH, A. D., KATZ, R., KONWINSKI, A., LEE, G., PATTERSON, D., RABKIN, A. & STOICA, I. 2010. A view of cloud computing. *Communications of the ACM,* 53**,** 50-58.

AVISON, D. & FITZGERALD, G. 2003. *Information systems development: methodologies, techniques and tools*, McGraw Hill.

BENGURIA, G., ELVESÆTER, B. & ILIEVA, S. 2013. Deliverable D2. 6 REMICS Handbook, Final Release.

BENNETT, K. 1995. Legacy Systems: Coping with Success. *IEEE Softw.,* 12**,** 19-23.

BEZEMER, C.-P. & ZAIDMAN, A. Multi-tenant SaaS applications: maintenance dream or nightmare?  Proceedings of the Joint ERCIM Workshop on Software Evolution (EVOL) and International Workshop on Principles of Software Evolution (IWPSE), 2010. ACM, 88-92.

BISBAL, J., LAWLESS, D., WU, B. & GRIMSON, J. 1999. Legacy information systems: Issues and directions. *Software, IEEE,* 16**,** 103-111.

BISBAL ,J., LAWLESS, D., WU, B., GRIMSON, J., WADE, V., RICHARDSON, R. & O'SULLIVAN, D. An overview of legacy information system migration.  Software Engineering Conference, 1997. Asia Pacific... and International Computer Science Conference 1997. APSEC'97 and ICSC'97. Proceedings, 1997. IEEE, 529-530.

BLODGET, H. 2011. Amazon's Cloud Crash Disaster Permanently Destroyed Many Customers' Data. *Available at* http://www.businessinsider.com.au/amazon-lost-data-2011-4 *[Last access June 2015*.[

BÖRNER, R. 2010. Applying situational method engineering to the development of service identification methods.

BRAUN, V. & CLARKE, V. 2006. Using thematic analysis in psychology. *Qualitative research in psychology,* 3**,** 77-101.

BREBNER, P. C. Is your cloud elastic enough?: performance modelling the elasticity of infrastructure as a service (IaaS) cloud applications.  Proceedings of the third joint WOSP/SIPEW international conference on Performance Engineering, 2012. ACM, 263-266.

BRINKKEMPER, S. 1996. Method engineering: engineering of information systems development methods and tools. *Information and software technology,* 38**,** 275-280.

BRODIE, M. & STONEBRAKER, M. 1995. *Migrating legacy systems: gateways, interfaces \& the incremental approach*, Morgan Kaufmann Publishers Inc.

BUYYA ,R., YEO, C. S., VENUGOPAL, S., BROBERG, J. & BRANDIC, I. 2009. Cloud computing and emerging IT platforms: Vision, hype, and reality for delivering computing as the 5th utility. *Future Generation computer systems,* 25**,** 599-616.

CHAUHAN, M. A. & BABAR, M. A .Towards Process Support for Migrating Applications to Cloud Computing.  Cloud and Service Computing (CSC), 2012 International Conference on, 22-24 Nov. 2012 2012. 80-87.

CHITCHYAN, R., RASHID, A., SAWYER, P., GARCIA, A., ALARCON, M. P., BAKKER, J., TEKINERDOGAN, B., CLARKE, S. & JACKSON, A. 2005. Survey of aspect-oriented analysis and design approaches.

DALHEIMER, M. & PFREUNDT, F.-J. Genlm: license management for grid and cloud computing environments.  Cluster Computing and the Grid, 2009. CCGRID'09. 9th IEEE/ACM International Symposium on, 2009. IEEE, 132-139.

DEDEKE, A. 2012. Improving Legacy-System Sustainability: A Systematic Approach. *IT Professional,* 14**,** 38-43.

DIESTE, O. & PADUA, O. Developing search strategies for detecting relevant experiments for systematic reviews.  Empirical Software Engineering and Measurement, 2007. ESEM 2007. First International Symposium on, 2007. IEEE, 215-224.

ERLIKH, L. 2000. Leveraging legacy system dollars for e-business. *IT Professional,* 2**,** 17-23.

FAHMIDEH, M. 2015 .A Generic Process Metamodel For Cloud Migration. *Available at* https://www.dropbox.com/s/4ibnnl8l0drc66c/A%20Development%20of%20Metamodel%20for%20Cloud%20Migration%20Process.pdf?dl=0.

ZIMMERMANN, O., MIKSOVIC, C. & KÜSTER, J. M. 2012. Reference architecture, metamodel, and modeling principles for architectural knowledge management in information technology services. *Journal of Systems and Software,* 85**,** 2014-2033.

**Appendix A**
**Conducting Literature Review**

**PLANNING REVIEW**

**Step 1 Pilot Review and Defining Search Strings.** From the first pilot review, we discerned that authors do not necessarily use the terms process models, life cycle, or methodology to name their proposal. This was why our initial search strings missed some well-known paper related to cloud migration. To alleviate this issue, the search strings were refined based on the recommended guidelines described in (Dieste and Padua, 2007) in order to identify all relevant studies even though they could not be labeled as a pure software development methodology. The five steps were followed: 1) Defining main terms by decomposing the research questions; 2) Identifying alternative synonyms for the main terms; 3) Checking the search strings in any relevant papers that retrieved; 4) Incorporating alternative synonyms using the logical operator 'OR'; and 5) Using the logical operator 'AND' to link the main terms. The terms "Cloud", "Cloud Computing", "Service Computing", "Legacy", "Methodology", "Process Model", "Reference Model", "Migration", "Framework" were set as the main keywords and based upon them, the different search strings were defined using the logical operator OR to include synonyms for each search string as well as the logical operator AND to link together each set of synonyms. Using the appropriate Boolean expressions, a set of search strings were generated as shown in Table III.

| Table III. List of search strings |
|---|
| Search Query (SQ) |
| **SQ1:** "Migration" OR "Cloud adoption" OR "Cloud migration" OR "migration to Cloud" OR "Legacy to Cloud migration" OR "Legacy migration to Cloud" AND [SQ2 OR SQ3] |
| **SQ2:** "Monolith application" OR "Legacy code" OR "Legacy system" OR "Legacy information systems" OR "Existing system" OR "Legacy component" OR "Legacy software" OR "Legacy application" "On-premise application" OR "Monolithic system" OR "Existing software" OR "Pre-existing software" OR "Legacy information system" OR "Legacy program" OR "Pre-existing assets" OR "Legacy architecture" OR "Legacy asset" |
| **SQ3:** "Methodology" OR "Software Development Methodology" OR "Development Process" OR "Process Model" OR "Reference Model" OR "Migration" OR "Framework" |

**Step 2 Selecting Study Sources.** The following databases were searched against predefined search strings: IEEE Explore, ACM Digital Library, SpringerLink, ScienceDirect, Wiley InterScience, ISI Web of Knowledge, and Google scholar. These databases cover the vast majority of published studies in the software engineering field. In addition, journals, conference, workshop proceedings and technical reports attributed to Cloud Computing and SOA areas were sought. These included IEEE Transaction of Cloud Computing, IEEE Transaction of Service Computing, Software Engineering for Cloud Computing, Cloud Computing International Conference, Cloud and Service Computing International Conference, International Conference of Service-Oriented Computing, Maintenance and Evolution of Service-Oriented and Cloud-Based Systems, International Conference on Cloud, Service-Oriented Computing and Applications, and International Conference on High Performance Computing and Communications.

**Step 3 Defining Study Inclusion and Exclusion Criteria.** The criteria for selecting a study were as follows:
— Relevant to the first research question (RQ1) though it could be titled under different terms,
— Focused on the migration of legacy applications to cloud environments, and directly dealt with the challenges as stated in Section 2.3,
— Published between 2007 (the date of origination of cloud computing) and June 2015.

This survey focused on studies pertaining to approach for moving one or more tier of legacy applications to cloud environments through proposing a process model or framework. The following exclusion criteria were set for:

- High-level frameworks suggesting conceptual models for outsourcing business processes to cloud environments and do not directly deal with legacy application migration were considered out the scope of this research. Consequently, these frameworks were excluded: Business Model Framework (CBMF), Linthicum Cloud Computing Framework (LCCF), Oracle Consulting Cloud Computing Services Framework, IBM Framework for Cloud adoption (IFCA), BlueSky Cloud Framework for e-Learning, and Hybrid ITIL V3 Framework for Cloud.
- Decision making frameworks aiding users to rank, evaluation, and selection cloud providers that fit legacy application requirements.
- Studies in languages other than English were excluded from the review.

**CONDUCTING REVIEW**

**Step 1 Study Selection**. This step dealt with the seeking of the defined search strings (Step 1 of Planning) over scientific databases (Step 2 of Planning). A set of primary papers was identified as the result of the first iteration. Then the title, abstraction, and in the most cases, the content of each paper were scrutinised regarding the inclusion and exclusion criteria, as defined in Step 3 of Planning. Reference snowballing was conducted in the sense that studies were cited in the references or related work section of a paper, were fed into the conducting review process as new resources. In some cases, for example, study [S7] in Table Appendix B, that a paper was not online accessible, its main author was communicated in order to receive the paper. In the case of multiple publications from an author or a research community, the most recent and completed one was chosen for the review. Consequently, 43 papers were identified, as the output of this step, for the review after applying the inclusion and exclusion criteria.

**Step 2 Data Extraction from the Identified Studies**. Finally, the proposed criteria in the evaluation framework were used as a lens to extract key data from the reminder identified papers. These data items enabled us to capture the full details of the studies under review. The extracted data from the studies and their analysis are presented in Section 5.3 and 5.4. Apart from data extraction using the framework, data items pertaining to research quality were extracted for further assessment as discussed in Section 5.1. The criteria were borrowed from Critical Appraisal Skills Programme (Greenhalgh and Taylor, 1997) and those suggested conducting empirical research in software engineering by (Kitchenham et al., 2002). Appendix C presents these criteria.

**OVERVIEW OF APPROACHES**

**Distribution by year of publication.** Table IV presents the frequency of the proposed approaches since 2009. N shows the number of publications per year. For example, until 2009 there have been 4 publications. The highest publication rate is in 2013 with 17 papers (34.7%). As shown in this table there has been a continuous growth in publications from 2009 to 2013. The number of publications has decreased in years 2014 with 7 papers.

Table IV. Frequency of papers by year

| Year | N | % |
| --- | --- | --- |
| 2009 | 3 | 6.9 |
| 2010 | 3 | 6.9 |
| 2011 | 6 | 13.9 |
| 2012 | 8 | 18.6 |
| 2013 | 15 | 34.8 |
| 2014 | 7 | 16.2 |
| 2015 | 1 | 2.3 |
| Total | 43 | ~100 |

**Distribution by publication type.** Table V presents a classification of publications from academia and industrial sectors. It indicates that the most of contributions are related to conferences with 19 out of all 43 publications. The second ranked publication channel is journals with 15 publications in total. Among them, the journal of *Software Practice and Experience* published the most papers related to cloud migration approaches with 3 papers. White papers, workshops papers, book chapters, and dissertations are placed as third, fourth, and fifth regarding the number and percentage of the total publications, respectively.

Table V. Distribution by publication channel

| Channel Name | Publisher Name | N |
| --- | --- | --- |
| **Book Chapter** | | |
| International Conference on Cloud Computing and Services Science | Springer | 1 |
| Computer Communications and Networks | Springer | 1 |
| Cloud Computing | Springer | 1 |
| **Total:3   %6.1** | | |

**Journal**

| | | |
|---|---|---|
| Software Practice and Experience | Wiley Online Library | 3 |
| Cloud Computing with e-Science Applications | CRC Press / Taylor & Francis | 1 |
| International Journal of Big Data Intelligence | Perpetual Innovation Media | 1 |
| International Journal of Cooperative Information Systems | World Scientific | 1 |
| Journal of Software Maintenance and Evolution | Wiley InterScience | 1 |
| Journal of Systems and Software | ScienceDirect | 1 |
| International Journal of Cloud Computing and Services Science | Institute of Advanced Engineering and Science | 1 |
| International Journal of Mechatronics, Electrical and Computer Technology | - | 1 |
| IEEE Computer | IEEE | 1 |
| Computing | Springer | 1 |
| International Journal on Advances in Software | GEOMAR | 1 |
| IEEE Transaction on Cloud Computing | IEEE | 1 |
| Information Technology | De Gruyter | 1 |
| **Total:15   %30.6** | | |

**Conference**

| | | |
|---|---|---|
| Services | IEEE | 1 |
| Service-I | IEEE | 1 |
| International Conference on Cloud Computing and Service Computing | IEEE | 1 |
| International Conference on Internet and Web Applications and Services | ThinkMind | 1 |
| International Conference on e-Business Engineering | IEEE | 1 |
| International Conference on Object Oriented Programming Systems Languages and Applications | Unipub | 1 |
| International Conference on Cloud Computing | IEEE | 1 |
| International Symposium on Service-Oriented System Engineering | IEEE | 1 |
| International Conference on Service-Oriented and Cloud Computing | Springer | 1 |
| International Service-Oriented Computing and Applications | IEEE | 1 |
| International Conference on Cloud Engineering | IEEE | 1 |
| World Congress on Services | IEEE | 1 |
| International Conference on Computer Software and Applications | IEEE | 1 |
| International Conference on High Performance Computing | IEEE | 1 |
| International Conference on Cloud Computing | IEEE | 1 |
| International Conference on Cloud Computing, GRIDs, and Virtualization | XPS | 1 |
| International Working Conference on Software Architecture | ACM | 1 |
| Symposium on Symbolic and Numeric Algorithms for Scientific Computing | IEEE | 1 |
| IEEE International Symposium on the Maintenance and Evolution of Service-Oriented and Cloud-Based Systems | IEEE | 1 |

**Total:19   %38.8**

**Workshop**

| | | | |
|---|---|---|---|
| International Workshop on Software Engineering for Cloud Computing | ACM | | 1 |
| International Workshop on the Maintenance and Evolution of Service-Oriented and Cloud-Based Systems | IEEE | | 1 |
| International Workshop on Modeling in Software Engineering | IEEE | | 1 |
| International Workshop on Service-Oriented Application | Springer | | 1 |
| **Total:4   %8.2** | | | |
| **White Paper** | | | |
| Amazon, Cloud Standards Customer Council, | - | | 2 |
| **Total:5   %10.2** | | | |
| **Total all publication type: 43** | | | |

**Distribution by authors' nationality.** Table VI shows the geographical distribution of the identified papers. The value N in this table indicates the total number of times authors from a country published a paper on the topic of cloud migration. According to this table, German stands at the first place for publication in this field with proposing 9 approaches (18.4% of total publication). USA and Ireland stand in the second and third places with 6 and 5 publications, respectively. UK, Spain, and Denmark are with 2 published papers, following with Italy, Belgium, Bulgaria, Portugal, Norway, Greece, Sweden, Finland with 1 published papers. An ascending ordering of publications based on the continent reveals cloud migration approaches in Europe with N=28, North and South America with N=8, Asia with N=7, and Australia with N=2. No contribution was found from Africa in this survey. It should be noted that Cisco, IBM, Logicalis, Cloud Standards Customer Council, and Remics consortium were not considered in country-based classification as they come from industry and dispersed over the world.

Table VI. Frequency of papers by year

| Continent | Country | N | % |
|---|---|---|---|
| Europe | Germany | 9 | 18.4 |
| | Ireland | 5 | 10.2 |
| | UK | 2 | 4.1 |
| | Spain | 2 | 4.1 |
| | Denmark | 2 | 4.1 |
| | Italy | 1 | 2 |
| | Belgium | 1 | 2 |
| | Bulgaria | 1 | 2 |
| | Portugal | 1 | 2 |
| | Norway | 1 | 2 |
| | Greece | 1 | 2 |
| | Sweden | 1 | 2 |
| | Finland | 1 | 2 |
| | **Total** | **28** | **56.9** |
| North and South America | USA | 6 | 12.2 |
| | Brazil | 1 | 2 |
| | Canada | 1 | 2 |
| | **Total** | **8** | **16.2** |
| Asia | China | 2 | 4.1 |
| | India | 2 | 4.1 |
| | United Arab Emirate | 1 | 2 |
| | Korea | 1 | 2 |
|  | Iran | 1 | 2 |
|  | **Total** | **7** | **14.2** |
| Australia | Australia | 2 | 4.1 |
|  | **Total** | **2** | **4.1** |

**Appendix B**

This appendix presents the list of identified approaches from the literature for the purpose of this survey.

Table B1 Studies included in final review

| Study ID | Authors and title | Abbreviation | Channel | Source | Year | Affiliation |
|---|---|---|---|---|---|---|
| [S1] | Krasteva, I., S. Stavru, et al., "*Agile Model-Driven Modernization to the Service Cloud*" | ICIW | Conference | ThinkMind | 2013 | Rila Solutions EAD, Bulgaria |
| [S2] | Beserra, P. V., A. Camara, et al, "*Cloudstep: A step-by-step decision process to support legacy application migration to the cloud*" | MESOCA | Workshop | IEEE | 2012 | Brazil |
| [S3] | Anstett, T., Leymann, F., Mietzner, R., "*Towards BPEL in the Cloud: Exploiting Different Delivery Models for the Execution of Business Processes*" | Services - I | Conference | IEEE | 2009 | Institute of Architecture of Application Systems, University of Stuttgart, Germany |
| [S4] | Conway, G. and E. Curry, "*The IVI Cloud Computing Life Cycle*" | CCSS | Book Chapter | Springer | 2013 | Innovation Value Institute, National University of Ireland |
| [S5] | Tran, V., J. Keung, et al, "*Application migration to cloud: a taxonomy of critical factors*" | SECLOUD | Workshop | ACM | 2011 | CSE, University of New South Wales, Australia |
| [S6] | Khajeh-Hosseini, A., D. Greenwood, et al, "*The cloud adoption toolkit: supporting cloud adoption decisions in the enterprise*" | SPE | Journal | Wiley Online Library | 2012 | Cloud Computing Co-laboratory, School of Computer Science University of St Andrews, UK |
| [S7] | Binz, T., Breitenbücher, U., "*Migration of enterprise applications to the cloud*" | - | Journal | Information Technology | 2014 | University of Stuttgart |
| [S8] | Chauhan, M. A. and M. A. Babar, "*Towards Process Support for Migrating Applications to Cloud Computing*" | CSC | Conference | IEEE | 2012 | Software & Systems Group IT University of Copenhagen, Denmark |
| [S9] | Strauch, S., et al. "*Migrating eScience Applications to the Cloud: Methodology and Evaluation*" | - | Journal | CRC Press / Taylor & Francis | 2014 | Institute of Architecture of Application Systems, University of Stuttgart, Germany |
| [S10] | S. Strauch, V. A., D. Karastoyanova, F. Leymann, "*Migrating Enterprise Applications to the Cloud: Methodology and Evaluation*" | JBDI | Journal | Perpetual Innovation Media | 2014 | Institute of Architecture of Application Systems, University of Stuttgart, Germany |
| [S11] | Leymann, F., et al., "*Moving applications to the cloud: An approach based on application model enrichment*" | JCIS | Journal | World Scientific | 2011 | Institute of Architecture of Application Systems, University of Stuttgart, Germany |
| [S12] | Zhang, W., et al., "*Migrating legacy applications to the service Cloud*" | OOPSLA | Conference | Unipub | 2009 | SINTEF, Norway |
| [S13] | La and Kim 2009, "*A systematic process for developing high quality saas cloud services*" | - | Book Chapter | Springer | 2009 | Department of Computer Science Soongsil University, Korea |
| [S14] | Miranda, J., et al., "*Assisting Cloud Service Migration Using Software Adaptation Techniques*" | Cloud Computing | Conference | IEEE | 2013 | Dept. of Information Technology and Telematic Systems Engineering, University of Extremadura, Cáceres, Spain |
| [S15] | Guillén, J., et al., "*A service-oriented framework for developing cross cloud migratable software*" | JSS | Journal | ScienceDirect | 2013 | GloIn, Calle Azorín 2, Cáceres, Spain |
| [S16] | SUN, K., Li, Y., "*Effort Estimation in Cloud Migration Process*" | SOSE | Conference | IEEE | 2012 | IBM Research - China |
| [S17] | Rabetski, P. and G. Schneider, "*Migration of an On-Premise Application to the Cloud: Experience Report*" | ESOCC | Conference | Springer | 2013 | Department of Computer Science and Engineering Chalmers University of Technology, and the University of Gothenburg Gothenburg, Sweden |
| [S18] | C., Pahl, H. Xiong, et al., "*A Comparison of On-Premise to Cloud Migration Approaches*" | SOCC | Conference | Springer | 2013 | IC4, Dublin City University, Dublin, Ireland |
| [S19] | Laszewski, T. and P. Nauduri, "*Migrating to the Cloud: Oracle Client/Server Modernization*" | Book | - | Elsevier | 2012 | USA |
| [S20] | Jamshidi, P., Pahl, C., "*Cloud Migration Patterns: A Multi-Cloud Architectural Perspective*" | WESOA | Workshop | Springer | 2014 | IC4 – the Irish Centre for Cloud Computing and Commerce, Dublin City University Dublin, Ireland |
| [S21] | Ahmad, Aakash, and Muhammad Ali Babar., "*A framework for architecture-driven migration of legacy systems to cloud-enabled software*" | WICSA | Conference | ACM | 2014 | IT University of Copenhagen, Denmark |
| [S22] | Ridha, Gadhgadhi, Khazri Saida, and Cheriet Mohamed. "*OPENICRA: Towards A Generic Model* | IJ-CLOSER | Journal | Institute of | 2013 | Multimedia Communication in Telepresence, Montréal |

| Ref | Citation | Venue | Type | Publisher | Year | Affiliation |
|---|---|---|---|---|---|---|
| | *for Automatic Deployment of Applications in the Cloud Computing"* | | | Advanced Engineering and Science | | (QC), Canada |
| [S23] | Council, C. S. C., *"Migration applications to public Cloud Services: roadmap for success"* | - | - | Cloud Standards Customer Council | 2013 | Cloud Standards Customer Council |
| [S24] | Ahmad Jalili, and Omid Bushehrian, *"A Structured and Achromatic Process for Legacy to Cloud Migration: A Survey"* | IJMEC | Journal | - | 2014 | Department of Computer Engineering and Information Technology, Shiraz University of Technology, Shiraz, Iran |
| [S25] | Quang Hieu, V. and R. Asal, *"Legacy Application Migration to the Cloud: Practicability and Methodology"* | SERVICES | World Congress | IEEE | 2012 | ETISALAT BT Innovation Centre Khalifa University, UAE, United Arab Emirates |
| [S26] | Benguria,G., Elvesæter, B., Ilieva, S. *"REuse and Migration of legacy applications to Interoperable Cloud Services"* | - | REMICS | - | 2013 | SINTEF ICT |
| [S27] | S., Rajaraajeswari, R., Pethuru, *"Cloud Application Modernization and Migration Methodology"* | - | Book Chapter | Springer | 2013 | Department of Master of Computer Applications, India |
| [S28] | Tang,K., Zhang J.M, Feng, C.H, *"Application Centric Lifecycle Framework in Cloud"* | E-Business Engineering | Conference | IEEE | 2011 | IBM Research China |
| [S29] | Maenhaut, p., Moens, H., Ongenae, V., Turck, F *"Migrating legacy software to the cloud: approach and verification by means of two medical software use cases"* | SPE | Journal | Wiley Online Library | 2015 | Ghent University, Belgium |
| [S30] | Binz, Leymann et al., *"CMotion: A framework for migration of applications into and between clouds"* | SOCA | Conference | IEEE | 2011 | Institute of Architecture of Application Systems University of Stuttgart, Germany |
| [S31] | Jamshidi, P., Ahmad, A., Pahl, C., *"Cloud Migration Research: A Systematic Review"* | TCC | Journal | IEEE | 2013 | School of Computing, Dublin City University, Ireland |
| [S32] | Ardagna, D., Casale., G., *"MODACLOUDS: A Model-Driven Approach for the Design and Execution of Applications on Multiple Clouds"* | MiSE | Conference | IEEE | 2012 | Politecnico di Milano |
| [S33] | Wu, J., et al., *"Migrating a Digital Library to a Private Cloud"* | ICE | Conference | - | 2014 | Information Sciences and Technology Computer Science and Engineering Pennsylvania State University, USA |
| [S34] | Pahl, C. and H. Xiong, *"Migration to PaaS clouds-Migration process and architectural concerns"* | MESOCA | Conference | IEEE | 2013 | the Irish Centre for Cloud Computing and Commerce, Ireland |
| [S35] | Andrikopoulos, V., Binz, T., Leymann, F., Strauch, S., *"How to Adapt Applications for the Cloud Environment"* | Computing | Journal | Springer | 2013 | Institute of Architecture of Application Systems, University of Stuttgart, Germany |
| [S36] | Bahga, A., Madisetti, V.K., *"Rapid Prototyping of Multitier Cloud-Based Services and Systems"* | IEEE Computer | Journal | IEEE | 2013 | Georgia Tech, USA |
| [S37] | C. Ward, N. Aravamudan, *"Workload Migration into Clouds – Challenges, Experiences, Opportunities"* | Cloud | Conference | IEEE | 2010 | IBM |
| [S38] | Lindner, M.A., McDonald, F., Conway, G., Curry E., *"Understanding Cloud Requirements - A Supply Chain Lifecycle Approach"* | - | Conference | XPS | 2011 | SAP Research, Palo Alto, USA |
| [S39] | Zhou, H., Yang, H.,Hugill, A., *"An Ontology-Based Approach to Reengineering Enterprise Software for Cloud Computing"* | COMPSAC | Conference | IEEE | 2010 | Software Technology Research Laboratory De Montfort University Leicester, UK |
| [S40] | Varia, J., *"Migrating Your Existing Application to the AWS Cloud. A Phase-Driven Approach to Cloud Migration"* | - | White paper | Amazon | 2010 | Amazon, USA |
| [S41] | Frey, S. and W. Hasselbring, *"The cloudmig approach: Model-based migration of software systems to cloud-optimized applications"* | - | Journal | GEOMAR | 2011 | Software Engineering Group University of Kiel, Germany |
| [S42] | Banerjee, J., *"Moving to the cloud: Workload migration techniques and approaches"* | HiPC | Conference | IEEE | 2012 | IBM Kolkata, India |
| [S43] | Menychtas, A., Santzaridou,C., Kousiouris, G., Varvarigou, T., *"ARTIST Methodology and Framework: A novel approach for the migration of legacy software on the Cloud"* | SYNASC | Conference | IEEE | 2013 | National Technical University of Athens Athens, Greece |

**Appendix C**

Table C1. Research quality criteria (Adapted from (Greenhalgh and Taylor, 1997) and (Kitchenham et al., 2002))

| Criteria Name | Definition | Criterion Type | Possible Values |
|---|---|---|---|
| Research Aim | Is there a clear statement of research objective? | Scale Form | ● The research provides a clear statement of the research objective.<br>⊖ The research provides a partial statement of the research objective.<br>○ The research does not provide a statement of the research objective. |
| Research Context | Is there an adequate description of the context in which the research was conducted? | Scale Form | ● The research clearly describes the research context.<br>⊖ The research partially describes the research context.<br>○ The research does not describe the research objective. |
| Research Design | Is there any appropriate research design to conduct the research? | Scale Form | ● The research describes the research design.<br>⊖ The research provides a partial description of the research design.<br>○ The research does not describe the research design. |
| Data Collection | Is there any clear description of data collection method to address research question? | Scale Form | ● The research provides a clear description of data collection method.<br>⊖ The research provides a partial description of the data collection method.<br>○ The research does not provide a description of the data collection method. |
| Data Analysis | Is there a clear and rigors description of data analysis? | Scale Form | ● The research clearly describes the data analysis.<br>⊖ The research partially describes the data analysis.<br>○ The research does not describe the data analysis. |
| Reflexivity | Is there an appropriate relationship between research and participants to conduct research? | Scale Form | ● The research clearly describes the data analysis.<br>⊖ The research partially describes the data analysis.<br>○ The research does not describe the data analysis. |
| Finding | Is there a clear description of research findings? | Scale Form | ● The research provides a clear description of the research findings.<br>⊖ The research provides a partial description of the research |

| | | | findings.<br>○ The research does not provide a description of the research finding. |
|---|---|---|---|
| Value | Is there a clear description of contributions to research and practice? | Scale Form | ● The research provides a clear description of the research contribution.<br>⊖ The research provides a partial description of the research contribution.<br>○ The research does not provide a description of the research contribution. |
| Validation Type | Does a validation of the suggested methodology exist? | Descriptive | - |

**Appendix D**

Table D.1. Description of included cloud migration approaches in final review

| Year | Study ID | Aim |
|------|----------|-----|
| 2009 | [S12] | The methodology proposes seven generic activities to help developers to move monolithic legacy applications to SaaS. |
| 2009 | [S13] | As conventional approaches such as object-oriented methodologies lack effectively support SaaS-specific engineering activities such as modelling common features, variability, and designing quality services, this paper presents a systematic process model for developing SaaS applications. It uses product-line engineering notions such as commonality and variability (C&V) modelling to maximise the reusability |
| 2009 | [S3] | This paper investigates how legacy business process models can be executed in the cloud using service delivery models such as IaaS, PaaS, and SaaS. It also describes modifications are required to execute BPEL processes in the cloud. The focus is on moving business logic of legacy assets to the cloud whilst data are kept in the local data center. |
| 2010 | [S39] | This paper proposes a five-step ontology development process to re-engineer legacy enterprise applications for shifting to the cloud by building an ontology for enterprise software and then partitioning and decomposing it into potential service candidates. The approach helps to understand legacy software and identify potential service candidates that can meet the objectives of cloud migration. |
| 2010 | [S40] | A phase-driven step-by-step methodology for moving enterprise applications to Amazon cloud. The methodology presents several scenarios of deploying applications in Amazon. |
| 2010 | [S37] | This paper proposes a framework, called Darwin, with an emphasis on Accelerating heterogeneous source/target migrations to standard virtualised environments. The model helps for smooth migration of the legacy workload from the local environment to the new cloud enabled environment in a cost effective way, with minimal disruption and risk. |
| 2011 | [S5] | The authors have a cost-oriented view to migration process and based on this propose a taxonomy of the migration activities. They show a breakdown of costs and factors that impact on activities in a case-study in a .NET n-tier application is migrated to run on Windows Azure. This taxonomy and cost factors contributed towards a cost estimation model of cloud migration. |
| 2011 | [S11] | A methodology and corresponding tool chains to rearrange the legacy application components and classify them into groups of components so that each group can be provisioned separately to different clouds while preserving the desired properties of the whole application. The methodology has been described in terms of the various activities to be performed and artefacts to be created in order to move an application to the cloud. |
| 2011 | [S28] | An application centric lifecycle management framework to deploy a complex application in the cloud. This framework exploits a model-driven approach which captures information in models that can be used to automatically generate code and configuration. |
| 2011 | [S30] | This paper proposes a framework called CMotion to address the vendor lock-in problem of application migration into and between different clouds. It uses adaptors to make previously incompatible technologies able to work together. Proposing various adaptation techniques is the main idea behind this framework. |
| 2011 | [S38] | The authors define a cloud lifecycle process from the supply chain management perspective and demonstrate how the supply chain plays an active role to manage the process for migrating from a traditional computing environment to a cloud hosting environment |
| 2011 | [S41] | A model-driven methodology including a set of reengineering activities in order to translate legacy application architecture to a target architecture for IaaS and PaaS-based cloud environments. |
| 2012 | [S2] | This paper presents Cloudstep, a step-by-step process from decision making viewpoint. It helps developers not only in selecting the cloud models and services best suited for their application, but also in carefully assessing the various risks and benefits involved. |
| 2012 | [S6] | A cloud adoption conceptual framework is proposed to support decision makers in identifying uncertainty to deploy legacy applications in the cloud and analysing the cost of deployment options which can affect bandwidth and the cloud service provider's pricing scheme. |
| 2012 | [S8] | This paper proposes a process model on the basis of gained experience in moving an Open Source System (OSS), Hackystat, to two different cloud computing platforms. It provides guidelines, observations and lessons learned regarding key issues involved in the migration such as analysis of target cloud environments against specific application's requirements, evaluation of platforms for cloud specific quality attributes of SaaS applications, the impact of a target cloud platform on each of the migration activities, and the potential influence of different services offered by cloud environments on migration activities and on the new architecture of a system to be migrated. |
| 2012 | [S16] | A process-based effort estimation approach is presented to assess the investment of a cloud migration process before it is undertaken with a focus on infrastructure-level migration. |
| 2012 | [S19] | The waterfall life cycle model has been used as a base to define a new methodology, concerning with activities such assessment, selecting service provider, migration effort estimation, training requirements, IT resource requirement for the migration project, capturing exhaustive amounts of information from the legacy systems, and database schema layout, testing, |

| Year | Ref | Description |
|------|-----|-------------|
| | | optimisation, and deployment. |
| 2012 | [S24] | A generic process model for migration with a focus on tools and techniques which can be applied in every step. |
| 2012 | [S25] | A methodology to assess the feasibility of migration to PaaS along with compatibility checklist to estimate the cost of migration as well as general solutions to address incompatibilities. |
| 2012 | [S42] | This paper offers a five-step migration methodology without re-architecting or re-engineering the existing applications. It also includes a set of high-level migration patterns emerges from the commonly found repeatable migration scenarios. |
| 2012 | [S32] | A model-driven approach with a focus on interoperability of applications in multiple cloud environments. In this approach, application models are designed at a high level to abstract from the targeted Cloud, and then semi-automatically translated into to run on multi-Cloud platforms. |
| 2013 | [S1] | This paper presents how model-driven modernisation can be enriched with Agile software development practices for seamless execution of different migration activities. |
| 2013 | [S4] | A nine-step cloud life cycle which is concerned with challenges such as security, data ownership, interoperability, service maturity and return on investment. It includes the recommended activities outputs to address these concerns. |
| 2013 | [S14] | An adaptation process to resolve incompatibility and vendor lock-in issues between legacy applications and cloud services. The approach promotes the use of decoupling-aware mechanisms in the design of cloud applications. |
| 2013 | [S15] | A development processes for developing cloud agnostic applications that may be deployed indifferently across multiple cloud platforms. In this approach information about cloud deployment and cloud integration is separated from the source code and managed by the framework. Interoperability between interdependent components deployed in different clouds is achieved by automatically generating services and service clients. |
| 2013 | [S17] | This paper presents an experience of a migration process to Microsoft Azure platform in the context of a small company. It recommends that activities and consideration should be taken into account in different migration scenarios, in particular from a performance and cost perspective. |
| 2013 | [S18] | The aim is to understand the core elements of cloud migration processes from an architecture perspective. Authors extract commonalities and difference between migration process activities for migration to IaaS, SaaS, and PaaS and present a range of concerns such architecture settings, costs, skills and technologies. |
| 2013 | [S22] | This paper focuses on the design and the implementation of a new generic model for automatic application deployment, called OpenICRA. It aims to reduce application development complexity and to simplify cloud services deployment process. |
| 2013 | [S23] | The methodology provides a series of activities that developers should take into account to ensure successful migration of existing applications to IaaS. It details activities related to assessing applications and workloads, developing a business case and technical approach, addressing security and privacy requirements as well as managing migration process. |
| 2013 | [S34] | A generic architecture-centric process for migrating to PaaS. The focus is on programming and architecture design principles that need to be incorporated in a migration process to PaaS. Experts' opinions from two case studies have been used to formulate this process model. |
| 2013 | [S36] | A three-phased architecture-centric development process for moving (or developing) multitier applications to the cloud with a focus on architecture design principles such as scalability, performance, maintainability, portability, and interoperability. |
| 2013 | [S43] | The methodology is a model-driven migration methodology and considers both the technical and business aspects of the legacy applications. The methodology emphasises generating reusable artefacts which can be transformed to multiple cloud platforms and facilitating automation of the migration. |
| 2013 | [S35] | This paper categorises different migration types and identifies the potential impact and adaptation needs for each of these types on the application tiers. It also investigates various cross-cutting concerns such as security that need to be considered for the migration of the application, and position them with respect to the identified migration types. |
| 2013 | [S27] | A five-step process to re-discover the optimal balance of performance, agility, sustainability, and cost. |
| 2013 | [S31] | Through using systematic literature review, this paper introduces a conceptual process model called Cloud-RMM (Cloud - Reference Migration Model), which classifies key process areas related to cloud migration. |
| 2013 | [S26] | The main objective of the REMICS (REuse and Migration of legacy applications to Interoperable Cloud Services) project is to specify, develop and evaluate a tool-supported model-driven methodology for migrating legacy applications to interoperable Service Cloud platforms. |
| 2014 | [S33] | This paper presents activities that authors carried out to move CiteSeerX into a private cloud. It also reports a number of lesson learned from this migration experience, challenges encountered prior to and during the migration and post-migration issues and possible solutions. |

| Year | Ref | Description |
|---|---|---|
| 2014 | [S9] | This methodology consists of seven phases for moving the database tier of e-science applications to the cloud. It incorporates guidance on choosing a cloud database solution and required refactoring activities. |
| 2014 | [S10] | A step-by-step methodology for the migration of the database layer to the cloud and the refactoring of the application architecture. |
| 2014 | [S21] | An architecture-centric process, named Legacy-to-Cloud Migration Horseshoe, which views migration as a set of recurring problems that can be formulated as migration process patterns. |
| 2014 | [S20] | This paper proposes 15 fine-grained patterns for migration that target multi-cloud settings and are specified with architectural notations. |
| 2014 | [S7] | An approach which provides activities to maintain an overall picture of the enterprise application as well as its environment and how to deploy and manage complex enterprise applications. |
| 2015 | [S29] | A generic model to migrate to a hybrid or public cloud environment, and the steps required to add multi-tenancy to these applications. |

**Appendix E**

Table E1. Result of evaluation of approaches based on generic criteria process clarity, procedure and technique, modelling, tailorability, and tool support

| Study | Process Clarity | Procedures & Techniques | Modelling | Tailorability | Tool Support |
|---|---|---|---|---|---|
| [S1] | ◐ | ○ | ◐ | ○ | ○ |
| [S2] | ◐ | ○ | ○ | ○ | ○ |
| [S3] | ◐ | ● | ○ | ○ | ○ |
| [S4] | ● | ● | ○ | ○ | ○ |
| [S5] | ● | ◐ | ○ | ○ | ○ |
| [S6] | ● | ● | ○ | ○ | ◐ |
| [S7] | ◐ | ◐ | ○ | ○ | ○ |
| [S8] | ● | ● | ○ | ○ | ○ |
| [S9] | ● | ● | ○ | ◐ | ◐ |
| [S10] | ● | ● | ○ | ◐ | ◐ |
| [S11] | ● | ● | ◐ | ○ | ◐ |
| [S12] | ● | ● | ◐ | ○ | ○ |
| [S13] | ● | ● | ◐ | ○ | ○ |
| [S14] | ● | ● | ○ | ○ | ○ |
| [S15] | ● | ● | ◐ | ○ | ◐ |
| [S16] | ● | ● | ○ | ○ | ○ |
| [S17] | ● | ● | ◐ | ○ | ○ |
| [S18] | ● | ◐ | ○ | ○ | ○ |
| [S19] | ● | ● | ○ | ○ | ○ |
| [S20] | ● | ● | ◐ | ◐ | ○ |
| [S21] | ● | ◐ | ○ | ○ | ○ |
| [S22] | ● | ◐ | ○ | ○ | ○ |
| [S23] | ● | ● | ○ | ○ | ○ |
| [S24] | ● | ○ | ○ | ○ | ○ |
| [S25] | ● | ◐ | ○ | ○ | ○ |

| | C1 | C2 | C3 | C4 | C5 |
|---|---|---|---|---|---|
| [S26] | ● | ● | ● | ◐ | ● |
| [S27] | ● | ◐ | ○ | ○ | ○ |
| [S28] | ● | ○ | ○ | ○ | ○ |
| [S29] | ● | ● | ◐ | ○ | ○ |
| [S30] | ● | ◐ | ○ | ○ | ○ |
| [S31] | ○ | ○ | ○ | ○ | ○ |
| [S32] | ◐ | ○ | ○ | ○ | ○ |
| [S33] | ◐ | ○ | ○ | ○ | ○ |
| [S34] | ● | ◐ | ○ | ○ | ○ |
| [S35] | ◐ | ◐ | ○ | ○ | ○ |
| [S36] | ● | ◐ | ◐ | ○ | ○ |
| [S37] | ● | ◐ | ◐ | ○ | ○ |
| [S38] | ● | ◐ | ○ | ○ | ○ |
| [S39] | ● | ● | ◐ | ○ | ◐ |
| [S40] | ● | ● | ◐ | ○ | ◐ |
| [S41] | ● | ● | ○ | ○ | ○ |
| [S42] | ● | ● | ○ | ○ | ○ |
| [S43] | ● | ● | ● | ◐ | ● |
| Fully-Supported | 35 (81%) | 23 (53%) | 2 (5%) | 0 | 2 (5%) |
| Partially-Supported | 7 (16%) | 14 (32%) | 12 (28%) | 5 (11%) | 7 (16%) |
| Not-Supported | 1 (2%) | 6 (14%) | 29 (67%) | 38 (88%) | 34 (79%) |
| Total | 43 | 43 | 43 | 43 | 43 |

**Appendix F**

Table F1 Generic criteria

| Criterion | Evaluation question | Criterion Type | Description | Source |
|---|---|---|---|---|
| Process clarity | Does the approach provide a clear description of the suggested phases and activities? | Scale | ● The approach explicitly provides a clear description of the activities for conducting the migration process.<br>⊖ The approach provides a general description for some activities but details are lacking.<br>○ The approach provides either very partial definition or any definition for the activities. | (Ramsin and Paige, 2008) |
| Procedures & Supportive Techniques | (i) what are the techniques to perform each activity?<br>(ii) are examples and heuristics of the activities provided? | Scale | ● The approach offers techniques/example for activities.<br>⊖ The approach offers techniques/example for some activities.<br>○ The approach does not provide supportive techniques or example for activities. | (Karam and Casselman, 1993) |
| Modeling Language | Does the approach specify a modelling or notational component to represent produced work-products during migration process? | Scale | ● An existing modeling language or a new one is prescribed for all of the activities.<br>⊖ An existing modeling language or a new one is prescribed for some of the activities.<br>○ Any modeling language has not been specified. | (Karam and Casselman, 1993);(Sturm and Shehory, 2004); |
| Traceability | Does the approach determine the sequence of modelling or dependencies between produced work-products? | | ● The approach specific traceability of all work-products in the migration process.<br>⊖ The approach specific traceability for a subset of activities.<br>○ Traceability links between work-products have not been specified. | (Karam and Casselman, 1993);(Ramsin and Paige, 2008);(Chitchyan et al., 2005) |
| Tailorability | (i) is the approach based on a one-fits-all assumption or define adaptation mechanisms for migration project?<br>(ii) is the approach components have been expressed in the form of method fragments or process components? | Scale | ● The approach defines mechanisms to configure and modify its suggested process or modeling language.<br>⊖ The approach provides a basis (e.g. repository of method fragments) so that the tailoring process is facilitated.<br>○ Any tailoring is supported by the approach. | (Karam and Casselman, 1993);(Ramsin and Paige, 2008) |
| Tool Support | (i) has the approach developed specific tools to perform migration process?<br>(ii) does the approach use or provide guidelines for developers to select existing third-party tools for modeling? | Scale | ● The approach provides tools for the whole of activities or integrated with existing third-party tools.<br>⊖ The approach provides tools for some activities but is lacking for other.<br>○ The approach neither provides tool nor refers to existing available third-party tools for modeling work-products. | (Karam and Casselman, 1993) |
| Theoretical | Is the approach has been inspired or developed based on | Descriptive | - | (Karam and |
| Foundation | the existing software engineering paradigms or practice? | | | Casselman, 1993) |
| Work-Products | What work-products are prescribed by the approach to produce in a migration process? | Descriptive | - | (Karam and Casselman, 1993);(Sturm and Shehory, 2004); |
| Development Roles | What development roles, who are responsible for performing migration activities or any stakeholder who are involved, are defined by approach? | Descriptive | - | (Karam and Casselman, 1993);(Sturm and Shehory, 2004); |
| Domain Applicability | What are application domains for which the approach can be proper? | Descriptive | - | (Karam and Casselman, 1993) |
| Scalability | Does the approach or a subset of it is applicable to handle various migration sizes? | | Yes: The approach explicitly defines mechanisms to support various migration sizes and its scalability has been demonstrated in real project. | (Sturm and Shehory, 2004);(Karam and Casselman, 1993);(Chitchyan et al., 2005) |
| | Does the methodology define mechanism or guidelines suitable to handle different migration sizes? | Boolean | No: The approach does not support scalability | |
| | Is there any real evidence of applying the approach in the migration of legacy application with different sizes? | | | |
| Formality | Does the approach provide a degree of formality on technical aspects? | | Yes: The approach provides formal techniques for some activities. No: Any formalism is supported by the approach. | (Karam and Casselman, 1993) |
| | Does the approach use unambiguous mathematical definitions for modeling or describing work-products? | Boolean | | |

**Appendix G**

Table G1. Evaluation results based on the cloud-specific criteria

| Study | Context Analysis | Legacy Application Understanding | Migration Requirement Analysis | Migration Planning | Cloud Service Provider Selection | Negotiation and Licensing | Cloud Architecture Model Definition | Enabling Elasticity | Enabling Multi-Tenancy | Code Refactoring | Data Adaptation | Developing Integrators | Training | Test | Continuous Integration | Environment Configuration | Continuous Monitoring |
|---|---|---|---|---|---|---|---|---|---|---|---|---|---|---|---|---|---|
| [S1] | ○ | ● | ● | ○ | ○ | ○ | ● | ○ | ○ | ○ | ○ | ○ | ○ | ● | ○ | ○ | ● |
| [S2] | ● | ● | ○ | ● | ● | ○ | ○ | ○ | ○ | ○ | ○ | ○ | ○ | ○ | ○ | ○ | ○ |
| [S3] | ○ | ○ | ○ | ○ | ○ | ○ | ⊖ | ○ | ⊖ | ● | ⊖ | ○ | ○ | ○ | ○ | ● | ○ |
| [S4] | ● | ○ | ● | ● | ● | ● | ● | ○ | ○ | ○ | ○ | ○ | ○ | ⊖ | ○ | ○ | ● |
| [S5] | ○ | ○ | ○ | ○ | ⊖ | ○ | ○ | ○ | ○ | ● | ● | ○ | ● | ● | ○ | ● | ○ |
| [S6] | ● | ○ | ○ | ○ | ○ | ○ | ○ | ○ | ○ | ○ | ○ | ○ | ○ | ○ | ○ | ○ | ○ |
| [S7] | ○ | ● | ○ | ○ | ○ | ○ | ⊖ | ○ | ○ | ⊖ | ⊖ | ⊖ | ○ | ⊖ | ○ | ⊖ | ○ |
| [S8] | ○ | ○ | ● | ○ | ● | ○ | ○ | ○ | ○ | ● | ⊖ | ○ | ○ | ○ | ○ | ○ | ⊖ |
| [S9] | ○ | ○ | ○ | ● | ● | ○ | ○ | ○ | ○ | ● | ● | ○ | ○ | ○ | ○ | ● | ○ |
| [S10] | ○ | ○ | ○ | ● | ● | ○ | ○ | ○ | ○ | ● | ● | ○ | ○ | ○ | ○ | ● | ○ |
| [S11] | ○ | ○ | ○ | ○ | ○ | ○ | ● | ○ | ○ | ○ | ○ | ○ | ○ | ○ | ○ | ○ | ○ |
| [S12] | ○ | ● | ○ | ○ | ○ | ○ | ● | ○ | ○ | ● | ○ | ● | ○ | ○ | ○ | ● | ○ |
| [S13] | ○ | ○ | ● | ○ | ○ | ○ | ● | ○ | ⊖ | ⊖ | ⊖ | ○ | ○ | ○ | ○ | ○ | ○ |
| [S14] | ○ | ○ | ○ | ○ | ○ | ○ | ○ | ○ | ○ | ● | ● | ● | ○ | ○ | ○ | ○ | ○ |
| [S15] | ○ | ○ | ○ | ○ | ○ | ○ | ⊖ | ○ | ○ | ● | ○ | ● | ○ | ● | ○ | ○ | ○ |
| [S16] | ○ | ● | ● | ● | ○ | ○ | ● | ○ | ○ | ○ | ○ | ○ | ● | ● | ○ | ○ | ○ |
| [S17] | ⊖ | ⊖ | ⊖ | ○ | ● | ○ | ● | ○ | ○ | ⊖ | ⊖ | ⊖ | ○ | ○ | ○ | ○ | ○ |
| [S18] | ○ | ○ | ● | ● | ○ | ● | ● | ○ | ○ | ○ | ⊖ | ⊖ | ● | ⊖ | ○ | ○ | ⊖ |
| [S19] | ○ | ○ | ● | ● | ⊖ | ○ | ○ | ○ | ○ | ○ | ● | ⊖ | ● | ○ | ○ | ● | ○ |
| [S20] | ⊖ | ○ | ⊖ | ○ | ○ | ○ | ● | ○ | ○ | ⊖ | ○ | ● | ○ | ○ | ○ | ⊖ | ● |
| [S21] | ● | ● | ● | ● | ⊖ | ○ | ○ | ○ | ○ | ⊖ | ○ | ○ | ○ | ○ | ○ | ○ | ○ |
| [S22] | ○ | ○ | ○ | ○ | ○ | ○ | ● | ● | ○ | ○ | ○ | ⊖ | ○ | ○ | ○ | ○ | ● |
| [S23] | ● | ● | ● | ⊖ | ○ | ● | ● | ○ | ⊖ | ○ | ○ | ● | ● | ● | ○ | ● | ● |
| [S24] | ⊖ | ● | ⊖ | ○ | ○ | ○ | ○ | ○ | ○ | ⊖ | ⊖ | ⊖ | ○ | ○ | ○ | ○ | ⊖ |

| | 1 | 2 | 3 | 4 | 5 | 6 | 7 | 8 | 9 | 10 | 11 | 12 | 13 | 14 | 15 | 16 | 17 |
|---|---|---|---|---|---|---|---|---|---|---|---|---|---|---|---|---|---|
| [S25] | ⊖ | ○ | ○ | ○ | ⊖ | ○ | ○ | ○ | ○ | ● | ● | ● | ○ | ○ | ○ | ○ | ○ |
| [S26] | ○ | ● | ● | ⊖ | ○ | ○ | ● | ⊖ | ○ | ⊖ | ○ | ○ | ○ | ⊖ | ○ | ○ | ● |
| [S27] | ● | ● | ○ | ● | ○ | ○ | ● | ○ | ⊖ | ○ | ○ | ○ | ● | ○ | ○ | ○ | ○ |
| [S28] | ○ | ○ | ● | ○ | ○ | ○ | ● | ○ | ○ | ○ | ○ | ○ | ○ | ○ | ○ | ● | ● |
| [S29] | ○ | ○ | ○ | ○ | ● | ⊖ | ⊖ | ⊖ | ● | ⊖ | ⊖ | ○ | ● | ○ | ○ | ○ | ○ |
| [S30] | ○ | ○ | ○ | ○ | ○ | ○ | ○ | ○ | ○ | ⊖ | ○ | ● | ○ | ○ | ○ | ○ | ○ |
| [S31] | ● | ⊖ | ● | ○ | ⊖ | ○ | ⊖ | ○ | ⊖ | ⊖ | ⊖ | ○ | ⊖ | ⊖ | ○ | ⊖ | ○ |
| [S32] | ○ | ⊖ | ○ | ○ | ⊖ | ○ | ⊖ | ○ | ○ | ○ | ○ | ● | ○ | ⊖ | ⊖ | ⊖ | ⊖ |
| [S33] | ⊖ | ⊖ | ● | ● | ○ | ○ | ● | ○ | ○ | ⊖ | ⊖ | ○ | ○ | ● | ● | ○ | ● |
| [S34] | ○ | ○ | ● | ⊖ | ○ | ○ | ● | ○ | ○ | ○ | ○ | ● | ● | ○ | ○ | ○ | ○ |
| [S35] | ⊖ | ⊖ | ⊖ | ○ | ⊖ | ⊖ | ● | ○ | ⊖ | ● | ● | ● | ○ | ⊖ | ○ | ○ | ○ |
| [S36] | ○ | ○ | ○ | ○ | ○ | ○ | ● | ○ | ○ | ○ | ○ | ○ | ● | ○ | ○ | ○ | ○ |
| [S37] | ⊖ | ● | ○ | ⊖ | ○ | ○ | ⊖ | ○ | ○ | ○ | ○ | ○ | ⊖ | ○ | ⊖ | ○ | ○ |
| [S38] | ○ | ○ | ● | ● | ● | ○ | ● | ○ | ○ | ○ | ○ | ○ | ○ | ○ | ○ | ○ | ● |
| [S39] | ○ | ● | ○ | ○ | ○ | ○ | ○ | ○ | ○ | ⊖ | ○ | ○ | ○ | ○ | ○ | ○ | ○ |

| | | | | | | | | | | | | | | | | | | |
|---|---|---|---|---|---|---|---|---|---|---|---|---|---|---|---|---|---|---|
| [S40] | | ● | ◐ | ○ | ◐ | ○ | ● | ● | ◐ | ○ | ◐ | ◐ | ◐ | ○ | ● | ○ | ● | ● |
| [S41] | | ○ | ● | ○ | ○ | ◐ | ○ | ● | ○ | ○ | ◐ | ◐ | ○ | ○ | ○ | ○ | ○ | ○ |
| [S42] | | ○ | ● | ○ | ● | ◐ | ○ | ○ | ○ | ○ | ◐ | ● | ◐ | ○ | ◐ | ○ | ● | ◐ |
| [S43] | | ● | ● | ○ | ○ | ○ | ◐ | ◐ | ○ | ○ | ● | ○ | ○ | ○ | ○ | ○ | ○ | ● |
| Fully-Supported | | 9 | 15 | 15 | 12 | 8 | 4 | 21 | 1 | 1 | 11 | 8 | 10 | 8 | 8 | 1 | 10 | 11 |
| Partially-Supported | | 7 | 6 | 7 | 5 | 9 | 3 | 8 | 3 | 6 | 15 | 12 | 8 | 1 | 9 | 1 | 5 | 5 |
| Not-Supported | | 27 | 22 | 26 | 26 | 26 | 36 | 14 | 39 | 36 | 17 | 23 | 25 | 34 | 26 | 41 | 28 | 27 |
| Total | | 43 | 43 | 43 | 43 | 43 | 43 | 43 | 43 | 43 | 43 | 43 | 43 | 43 | 43 | 43 | 43 | 43 |

**Appendix H**

Table H1 Cloud-specific criteria

| Criterion | Evaluation question | Criterion Type | Possible evaluation result |
|---|---|---|---|
| Migration Type | What kind of service delivery model and application tier are concerned with the approach? | Multiple | Types I, II, III, IV, V |
| Unit of Migration | What tiers of application has the approach been designed for? | Multiple | Whole Application Stack, User Interface Tier, Business Logic Tier, Data Tier |
| Context Analysis | Does the approach define activities related to assess the suitability of migration to cloud in terms of security, legality, cost, organizational change and user resistance? | Scale | ● The approach explicitly provides a clear description of the activities for conducting a context analysis. <br> ⊖ The approach provides a general description for context analysis activity but details are lacking. <br> ○ The approach provides either very partial definition or any definition for the context analysis. |
| Legacy Application Understanding | (i) Does the approach provide activities for recapturing the abstract representation of application architecture in terms of functionalities, data, constraints, structure of components, business and data tiers, design quality, complexity and coupling? <br><br> (ii) Does the approach identify and collect usage data of working legacy application and its resource utilisation model? | Scale | ● The approach explicitly provides a clear description of the activities to discover knowledge about the legacy application. <br> ⊖ The approach provides a general description for discovering knowledge about legacy application but details are lacking. <br> ○ The approach provides either very partial definition or any definition for the legacy application understanding. |
| Migration Requirements Analysis | Does the approach provide activities to acquire a set of requirements from multiple stakeholders about the target application according to the current configuration setting of the legacy application? Requirements such as computational requirements, servers, data storage and security, networking and response time, and elasticity requirements | Scale | ● The approach defines activities related to elicit and analysis of migration requirements. <br> ⊖ The approach offers general guideline for requirement analysis. <br> ○ Migration requirement analysis is not addressed in the approach. |
| Planning | (i) Does the approach provide support for correct and safe sequence of steps which guide the rest of migration process? <br><br> (ii) Does the approach provide guidelines for a proper roll-back plan to an in-house version of the legacy application? | Scale | ● The approach defines activities related to migration planning. <br> ⊖ The approach offers general guideline for planning. <br> ○ The approach does not support planning activity. |
| Cloud Service Provider Selection | (i) Does the approach provide activities, guidelines, or concerns that should be taken into account to choose a cloud provider that meet requirements?<br><br>(ii) Does the approach address licensing issues in the cloud environments? | Scale | ● The approach defines criteria or guidelines for selection cloud services are to be utilised by the legacy application.<br>⊖ The approach offers general guideline for cloud service selection.<br>○ The approach does not support the activity of cloud provider selection. |
| Training | Does the approach specify necessary skill or training requirements for developers and IT staff? | Scale | ● The approach explicitly defines an activity for developers and IT staff training.<br>⊖ The approach suggests considering training activities but does not define activities or guidelines for it.<br>○ The approach does not incorporate training into its suggested process model. |
| Component selection and distribution in the cloud | What are the activities or guidelines in the approach to assess and determine legacy components which are suitable to be migrated to the cloud? | Scale | ● The approach provides an explicit activity for selecting legacy components which are suitable for migration to the cloud and the way of arranging them in cloud platforms.<br>⊖ Support for component selection and distribution is implicit and confined to general advice.<br>○ The approach does not support this activity. |
| Incompatibility Resolution (Application Refactoring) | (i) Does the approach identify possible incompatibilities and possible solutions to resolve these incompatibilities?<br><br>(ii) What kind incompatibilities are concerned with the approach? | Scale | ● The approach defines activities to identify incompatibilities between legacy components and cloud services and suggests techniques to resolve them.<br>⊖ The approach concerns incompatibilities between legacy application and cloud services, however, does not define an explicit activity to identify and address these incompatibilities.<br>○ The approach does not address incompatibilities. |
| Enabling Multi-Tenancy | (i) Does the approach provide activities, techniques, or guidelines for detecting and handling faults which might incur in a tenant (tenant availability isolation)?<br><br>(ii) Does the approach include activities for identifying commonality and variability in the target domain? Any specific techniques are offered for application customisability on the basis of particular (application customisability)? | Scale | ● The approach defines activities to address multi-tenancy.<br>⊖ The approach concerns with multi-tenancy but there is not guideline to enable multi-tenancy in the legacy application. |
| | (iii) Does the approach provide mechanisms to protect tenants from each other in terms of access to data and security (security isolation)? | | ○ The approach does not address multi-tenancy. |
| | (iv) Does the approach provide support to protect tenant's performance from being affected by other tenant's behavior (tenant performance isolation)? | | |
| Enabling Application Elasticity | Is there any activity or guideline in the approach to define scaling rules, dynamic acquisition and release of cloud resources? | Scale | ● The approach defines activities to address elasticity. ◐ The approach concerns with multi-tenancy but there is not guideline to enable elasticity in the legacy application. ○ The approach does not address elasticity. |
| Test and Continuous Integration | (i) What kind of test activities are defined in the approach? (ii) Is there any systematic support to apply changes to application parts which are host in the cloud? | Scale | ● The approach defines a detailed activity to conduct test and integration. ◐ The approach provides general guidelines to conduct test and integration. ○ The approach does not cover test and integration. |
| Environment Configuration | How approach provide support for re-configuring the running environment of the application including reachability policies to resources and network, connection to storages, setting ports and firewalls, and load balancer? | Scale | ● The approach defines a detailed activity to prepare cloud environment to deploy the application. ◐ The approach provides general guidelines for configuring the cloud environment prior deploying application. ○ The approach does not support environment configuration. |
| Continuous Monitoring | Does the approach provide mechanisms for continuously monitoring application components utilising cloud services e.g. CPU, Memory, Disk I/O, and Network I/O? | Scale | ● The approach explicitly provides an activity for continuous application monitoring when running in the cloud. ◐ The approach provides general advice for application monitoring in the cloud. ○ The approach does not support application monitoring. |

**Appendix I**

Table I.1 Quality Assessment of the Studies

| Study | Research Aim | Research Context | Research Design | Data Collection | Data Analysis | Reflexivity | Findings | Value |
|---|---|---|---|---|---|---|---|---|
| [S1] | ◐ | ○ | ○ | ○ | ○ | ○ | ○ | ○ |
| [S2] | ● | ◐ | ○ | ○ | ○ | ○ | ● | ○ |
| [S3] | ● | ◐ | ○ | ○ | ○ | ○ | ● | ● |
| [S4] | ● | ● | ● | ● | ◐ | ● | ◐ | ● |
| [S5] | ● | ● | ◐ | ● | ○ | ○ | ● | ● |
| [S6] | ● | ● | ◐ | ○ | ○ | ○ | ● | ● |
| [S7] | ● | ○ | ○ | ○ | ○ | ○ | ○ | ○ |
| [S8] | ● | ● | ◐ | ○ | ○ | ○ | ● | ● |
| [S9] | ● | ● | ◐ | ● | ◐ | ○ | ● | ● |
| [S10] | ● | ● | ◐ | ● | ◐ | ○ | ● | ● |
| [S11] | ● | ● | ◐ | ◐ | ◐ | ○ | ● | ● |
| [S12] | ● | ● | ◐ | ○ | ○ | ○ | ○ | ○ |
| [S13] | ● | ● | ◐ | ○ | ○ | ○ | ○ | ○ |
| [S14] | ● | ● | ○ | ○ | ○ | ○ | ○ | ○ |
| [S15] | ● | ● | ● | ● | ● | ● | ● | ● |
| [S16] | ● | ● | ◐ | ○ | ○ | ○ | ◐ | ● |
| [S17] | ◐ | ● | ● | ◐ | ◐ | ○ | ● | ● |
| [S18] | ● | ● | ◐ | ○ | ○ | ○ | ● | ● |
| [S19] | ◐ | ● | ○ | ○ | ○ | ○ | ○ | ○ |
| [S20] | ● | ● | ● | ◐ | ● | ◐ | ◐ | ◐ |
| [S21] | ● | ○ | ○ | ○ | ○ | ○ | ○ | ○ |
| [S22] | ● | ● | ◐ | ◐ | ◐ | ○ | ◐ | ● |
| [S23] | ● | ○ | ○ | ○ | ○ | ○ | ○ | ○ |
| [S24] | ● | ○ | ○ | ○ | ○ | ○ | ◐ | ◐ |
| [S25] | ● | ● | ○ | ○ | ○ | ○ | ◐ | ◐ |
| [S26] | ● | ● | ○ | ○ | ○ | ○ | ○ | ● |
| [S27] | ● | ○ | ○ | ○ | ○ | ○ | ◐ | ○ |
| [S28] | ● | ◐ | ○ | ○ | ○ | ○ | ◐ | ◐ |
| [S29] | ● | ● | ◐ | ○ | ○ | ○ | ● | ◐ |
| [S30] | ● | ○ | ○ | ○ | ○ | ○ | ● | ◐ |
| [S31] | ● | ● | ● | ● | ● | ● | ● | ● |
| [S32] | ● | ● | ○ | ○ | ◐ | ○ | ◐ | ○ |
| [S33] | ◐ | ○ | ○ | ○ | ○ | ○ | ● | ○ |
| [S34] | ● | ○ | ○ | ◐ | ○ | ○ | ● | ◐ |
| [S35] | ● | ◐ | ○ | ○ | ○ | ○ | ● | ● |
| [S36] | ● | ◐ | ● | ● | ● | ○ | ◐ | ◐ |
| [S37] | ● | ● | ○ | ◐ | ◐ | ○ | ○ | ○ |
| [S38] | ● | ○ | ○ | ○ | ○ | ○ | ○ | ○ |
| [S39] | ● | ◐ | ○ | ○ | ○ | ○ | ○ | ○ |
| [S40] | ● | ● | ○ | ○ | ○ | ○ | ○ | ○ |
| [S41] | ● | ● | ○ | ○ | ○ | ○ | ○ | ● |
| [S42] | ● | ○ | ○ | ○ | ○ | ○ | ○ | ○ |

| [S43] | ● | ⊖ | ○ | ○ | ○ | ○ | ○ | ● |
|---|---|---|---|---|---|---|---|---|
| Fully-Supported | 39(90%) | 25 (58%) | 6 (14%) | 7 (16%) | 4 (9%) | 3 (7%) | 17 (39%) | 18 (41%) |
| Partially-Supported | 4 (9%) | 7 (16%) | 12 (28%) | 6 (14%) | 8 (18%) | 1 (2%) | 10 (23%) | 8(18%) |
| Not-Supported | 0 | 11 (26%) | 25 (58%) | 30 (70%) | 31 (72%) | 39 (90%) | 16 (37%) | 17 (40%) |
| Total | 43 | 43 | 43 | 43 | 43 | 43 | 43 | 43 |

***Mahdi Fahmideh*** is the assistant professor of Information Technology in the Faculty of Engineering and Information Sciences at University of Wollongong, Australia. He received a PhD degree in Information Systems from the Business School of University of New South Wales, Sydney, Australia. His research focuses on creating new-to-the-world artifacts that help organizations in adopting IT initiatives to tackle problems. His research outcome can be in the form of methodological approaches, conceptual models, decision-making frameworks, and software tools. His research interests lie at the intersection of cloud computing, data analytics, IoT, blockchain, and method engineering. Prior to joining the academia, Dr. Fahmideh has worked as a software developer in different industry sectors including accounting, insurance, defence, and publishing

***Farhad Daneshgar*** received his PhD in Information Systems from the University of Technology, Sydney Australia. He is a Senior Lecturer at the UNSW Business School, University of New South Wales, Sydney, Australia, and is an adjunct Professor at Bangkok University. Farhad is the creator of the Awareness Modeling Language, and has published extensively in the areas of Knowledge Management and Enterprise Systems, and was awarded twice for his Outstanding Research Article at the University of New South Wales. Farhad is a member of editorial board in five academic journals.

***Graham Low*** received the BE and PhD degrees from The University of Queensland. He is an emeritus professor of information systems in the School of Information Systems, Technology and Management at the University of New South Wales. His research program focuses on the implementation and adoption of new technologies. This can take the form of new/modified approaches/techniques for information systems development such as methodological approaches to agent-oriented information systems design and management of the information systems design and implementation process.

***Ghassan Beydoun*** received a degree in Computer Science and a Ph.D. degree in Knowledge Systems from the University of New South Wales. He is currently an Associate Professor at the School of Computing and Information Technology at the University of Wollongong. He is also Director of Software Design Science Research Centre. He has authored more than 100 papers for international journals and conferences. He investigates the best uses of ontologies in developing methodologies for distributed systems. His other research interests include multi-agent systems applications, ontologies and their applications, knowledge acquisition and disaster management.